\documentclass[aps,showpacs,superscriptaddress,nofootinbib]{revtex4-2}
\usepackage{graphicx}
\usepackage{bm}
\usepackage{amsmath}
\usepackage{mathrsfs}
\usepackage{ulem}
\usepackage{color}
\usepackage{booktabs}
\usepackage{tikz}
\usepackage[colorlinks, linkcolor=red,anchorcolor=green,citecolor=blue]{hyperref}
\usepackage{amsthm}
\usepackage{float}

\newtheorem*{AP}{Argument Principle}

\graphicspath{{./figs/}}

\begin{document}

\title{Normal mode analysis within relativistic  massive transport}

\author{Xin Lin}
\affiliation{Department of Physics, Fuzhou University, Fujian 350116, China}

\author{Qiu-Ze Sun}
\affiliation{Department of Physics, Tsinghua University, Beijing 100084, China}

\author{Xin-Hui Wu}
\affiliation{Department of Physics, Fuzhou University, Fujian 350116, China}

\author{Jin Hu}
\email{hu-j23@fzu.edu.cn}
\affiliation{Department of Physics, Fuzhou University, Fujian 350116, China}

\begin{abstract}

 In this paper, we address the normal mode analysis of the linearized Boltzmann equation for massive particles in the relaxation time approximation. We find that the sound channel and the heat channel are coupled in the massive case, which contrasts with their decoupling in the massless relaxation time approximation. In the limit of zero mass, our results smoothly reduce to the known case where the two channels decouple. 
  By utilizing the argument principle in complex analysis, we determine the existence condition for collective modes and find the onset transition behavior of collective modes previously observed in massless systems. We numerically determine the critical wave number $\kappa_c$ for the existence of each mode under various values of the scaled mass. Within the range of scaled masses considered, the critical wave numbers for the heat and shear channels increase with increasing scaled mass, while that of the sound channel exhibits a nonmonotonic dependence on the scaled mass. In addition, we analytically derive the dispersion relations for these collective modes in the long-wavelength limit. Notably, kinetic theory also incorporates collisionless damping effects known as Landau damping. We find that the branch cut structure responsible for Landau damping differs  from the massless case: Whereas the massless system features only two branch points, the massive system exhibits an infinite number of such points forming a continuous branch cut.
    
\end{abstract}

\maketitle

\section{INTRODUCTION}

Relativistic hydrodynamics, as a universal effective theory for large distance and time scales, plays a central role in the study of high-energy physics \cite{Romatschke:2017ejr} and cosmology \cite{McDonough:2020tqq}.  Notably, extensive research efforts and successful applications have been devoted to understanding the evolution of fireballs generated in relativistic heavy-ion collisions. Specifically, by applying relativistic hydrodynamics to analyze experimental data from facilities such as the BNL-RHIC and CERN-LHC, researchers have made significant progress in elucidating the fundamental properties of quark-gluon plasma (QGP) under extreme conditions of temperature and density \cite{Hama:2004rr,Huovinen:2006jp,Jeon:2015dfa,Bernhard:2019bmu,Auvinen:2020mpc,Nijs:2020roc,JETSCAPE:2020mzn,Parkkila:2021tqq,Czajka:2017wdo}. Interestingly, even in smaller collision systems, such as those involving proton-nucleus and proton-proton collisions \cite{CMS:2016fnw,ATLAS:2017hap} where the applicability of relativistic hydrodynamics was initially questioned — experimental results indicate that hydrodynamic models still provide a good description of these phenomena. This observation prompts questions regarding the emergence of hydrodynamic behavior in nonequilibrium systems, requiring a reevaluation of the conditions required for such collective dynamics to emerge.

Moreover, the successful application of hydrodynamics to heavy-ion collisions remains puzzling, as it implies that the initially far-from-equilibrium system undergoes an extremely rapid thermalization process, the physical mechanism of which is not yet fully understood. This necessitates the inclusion of effective thermalization mechanisms in phenomenological models of relativistic heavy-ion collisions, so that the system can quickly reach local thermal equilibrium. Such a rapid thermalization is crucial for accurately simulating the evolution of the QGP and reproducing experimental observations. Nevertheless, uncovering the underlying mechanism of this rapid thermalization remains a major scientific challenge. More broadly, understanding how systems far from equilibrium evolve toward thermalization is a question of fundamental and universal significance. Recently, some researchers have pointed out that the validity of hydrodynamics, in essence, means that the hydrodynamic constitutive relations are approximately satisfied \cite{Heller:2011ju}. Therefore, even if the system deviates from local equilibrium, this does not necessarily imply that hydrodynamics is inapplicable. This perspective effectively redefines our understanding of the Landau paradigm of hydrodynamics: Traditionally, hydrodynamics has been viewed as an effective theory built upon the assumption of local thermal equilibrium \cite{Landau:Fluid}. Later on, researchers found that hydrodynamics possesses an intrinsic mathematical structure reminiscent of attractors in complex systems, which may explain its broad applicability: The system tends to evolve toward a universal attractor, independent of the specific initial conditions \cite{Heller:2015dha}. Recently, similar attractorlike structures have been observed across a wide range of physical systems with various symmetries and basic settings within the framework of relativistic hydrodynamics. These findings offer new theoretical insights into why hydrodynamic models remain valid even on extremely short timescales. Consequently, current research efforts are increasingly directed toward understanding the intrinsic connection between attractor behavior and early-time dynamics
\cite{Strickland:2018ayk,Denicol:2020eij,Behtash:2017wqg,Romatschke:2017vte,Chattopadhyay:2019jqj,Kurkela:2019set,Chen:2024pez}.

The aforementioned studies primarily focus on the internal structure of hydrodynamics. In recent years, however, increasing attention has been devoted to exploring the connection between hydrodynamics and kinetic theory \cite{Romatschke:2015gic,Kurkela:2017xis}. For weakly coupled nonequilibrium systems, the primary theoretical framework is kinetic theory, exemplified by the Boltzmann equation. A deeper investigation into the intrinsic relationship between kinetic theory and hydrodynamics, including both their shared features and fundamental differences, holds promise for revealing the limitations of hydrodynamic descriptions. In other words, this line of research further clarifies the conditions under which hydrodynamics remains applicable. As a first step toward clarifying the intrinsic connection between hydrodynamics and kinetic theory, studies of two-point retarded correlation functions have recently attracted widespread research interest \cite{Romatschke:2015gic,Kurkela:2017xis,Bajec:2024jez,Brants:2024wrx}. The nonanalytic structures of these correlators often dictate the dynamical behavior of such systems during their late-stage evolution. Specifically, the pole structure of correlation functions encodes information about collective excitations, while branch cuts and nonhydrodynamic poles  are related to the application range of hydrodynamic descriptions. In the context of large \(N\) thermal \(\mathcal{N} = 4\) super Yang-Mills theory at large t'Hooft coupling, studies have shown that the associated correlators exhibit only poles or quasinormal modes \cite{Hartnoll:2005ju,Kovtun:2005ev,Grozdanov:2016vgg}. This observation has inspired analogous investigations into weakly coupled kinetic theory. For instance, Paul Romatschke analyzed the physical behavior of thermal correlators using  effective kinetic theory in the relaxation time approximation (RTA) \cite{Romatschke:2015gic} (see also \cite{Bajec:2024jez} including the calculation of cross-correlators), revealing the dominant role of hydrodynamic poles as long-lived modes in the first (physical) Riemann sheet. Conversely, a study  \cite{Kurkela:2017xis} suggests that branch cuts are dominant nonanalytic structures based on a model of momentum-dependent RTA; see also \cite{Brants:2024wrx}.  Other analytical calculations \cite{strain2010,Gavassino:2024rck} and numerical simulations \cite{Moore:2018mma,Ochsenfeld:2023wxz} support this latter conclusion regarding the dominance of branch cuts. The observed discrepancies among various studies can be traced back to differences in the microscopic interaction details, which may significantly influence the nonanalytic structures exhibited by the correlation functions \cite{Hu:2024tnn} 
It is important to recognize that both perspectives are valid, provided the discussion is restricted to their respective applicable scenarios. In fact, within the context of kinetic theory, scattering processes are classified into hard and soft interactions according to the form of interparticle forces, specifically, the differential scattering cross section or interaction potential. These different types of interactions give rise to distinct spectrum structures of the linearized collision operator, which in turn directly influence the physical properties of retarded correlation functions, see \cite{Hu:2024tnn,libo,dud} for more technical details.

In this work, we focus on a weakly coupled massive transport system and adopt the Boltzmann equation within a novel relaxation time approximation where the collision invariance is explicitly recovered (we shall come to the novel relaxation time approximation in the main text later). This approach provides a simple yet analytically tractable framework for describing the dynamical evolution from far-from-equilibrium to near-equilibrium states, and thus serves as a promising candidate for clarifying the intrinsic connection between hydrodynamics and kinetic theory. Then we perform a linear analysis and seek the normal mode solution to the linearized Boltzmann equation, from which we extract a nonanalytic structure that is  different from that of a massless particle system. This paper is organized as follows: In Sec.~\ref{linearized}, we provide a concise overview of the linearized kinetic equations.  In Sec.~\ref{model}, we construct a transport model within a novel relaxation time approximation. Section.~\ref{main} is devoted to our main analysis, where we give a detailed normal mode analysis and obtain two distinct sets of normal mode solutions.  Finally, a summary and outlook are given  in Sec.~\ref{outlook}. Additional relevant materials are included in the Appendix. 

Throughout this work, we employ natural units with \(k_B = c = \hbar = 1\). The metric tensor is defined as \(g^{\mu\nu} = \text{diag}(1, -1, -1, -1)\), and the projection tensor orthogonal to the four-velocity \(u^\mu\) is given by \(\Delta^{\mu\nu} \equiv g^{\mu\nu} - u^\mu u^\nu\). The transverse derivative orthogonal to \(u^\mu\) is expressed as \(\nabla_\mu \equiv \Delta_\mu^\nu \partial_\nu = (\delta_\mu^\nu - u_\mu u^\nu) \partial_\nu\). For brevity, the abbreviation dP denotes \(\int dP \equiv \frac{2}{(2\pi)^3} \int d^4p\, \theta(p^0) \delta(p^2 - m^2)\) and $m$ represents the particle mass. Additionally, we use the symmetric shorthand notation \(X^{(\mu\nu)} \equiv \frac{X^{\mu\nu} + X^{\nu\mu}}{2}\).

\section{LINEARIZED KINETIC EQUATION}
\label{linearized}

We start with the relativistic Boltzmann equation in the absence of external field,
\begin{align}\label{eq:Boltzmann}
    p^{\mu}\partial_{\mu}
    f\left(x,p\right)=C[f],
\end{align}
which describes the evolution of the single-particle distribution function in phase space $(x^\mu,p^\mu)$. The symbol \(\partial_{\mu}\) denotes the partial derivative with respect to spacetime coordinates. The collision term \(C[f]\) encapsulates the effects of local particle collisions on the evolution of the distribution function
\begin{align}
\label{ckl}
    C\left [ f \right ] \equiv \int dP'dP_{1}dP_{2}W_{p,p'\to p_{1},p_{2}}
\left ( f\left ( p_{1} \right )f\left ( p_{2} \right )
\tilde{f} \left ( p\right )\tilde{f} \left ( p'\right ) 
- f\left ( p \right )f\left ( p' \right )
\tilde{f} \left ( p_{1}\right )\tilde{f} \left ( p_{2}\right )\right ) ,
\end{align}
where $W_{p,p'\to p_{1},p_{2}}$ denotes the transition rate encoding the microscopic dynamic details. Here we consider only two-body elastic collisions and \(\tilde{f}=1\pm f\) according to the statistical characteristics of fermions or bosons, with the \(+\) sign corresponding to bosons and the \(-\) sign to fermions. In the classical limit, \(\tilde{f}\) reduces trivially to unity, and we shall confine our discussion to this case. When writing the collision kernel in the form of Eq.(\ref{ckl}), we have implicitly adhered to the principle of detailed balance expressed as \(W_{p,p'\to p_{1},p_{2}}=W_{p_{1},p_{2}\to p,p'}\). Note as an aside, Eq.(\ref{ckl}) can naturally be extended to the multicomponent case  detailed in  \cite{DeGroot:1980dk, Hu:2022vph}.

Although we have introduced approximations such as classical statistics and two-body elastic scattering, the relativistic Boltzmann equation remains a highly complicated nonlinear integro-differential equation, making its solution extremely challenging, both analytically and numerically. Analytical solutions of the nonlinear relativistic Boltzmann equation are known only in a few special cases \cite{Bazow:2015dha,Hu:2024utr}. To proceed in kinetic theory, a standard approach involves linearizing the nonlinear Boltzmann equation through an expansion about the local equilibrium state  \(f_{0}(x,p)\), retaining only first-order terms in the deviation function $\chi(x,p)$, i.e., \(f = f_{0}(1 + \chi)\). The local equilibrium distribution function follows from the collision invariance of the  Boltzmann equation  and  can be represented by
\begin{align}
    f_{0}\left ( x,p \right ) =\exp\left [ \alpha(x)-\beta^{\mu}(x)  p_{{\mu}} \right ],
\end{align}
where \(\beta^{\mu} = \frac{u^{\mu}}{T}\), \(\alpha= \frac{\mu}{T}\), \(\beta = \frac{1}{T}\) with \(T,\mu\) denoting the temperature and  chemical potential associated with the conserved particle number. Substituting the expression \(f = f_{0}(1 + \chi)\) into Eq.\eqref{ckl} gives the right-hand side (rhs) of the linearized Boltzmann equation
\begin{align}\label{rhs}
    \text{rhs}=-E_pf_{0}\left ( x,p \right ) \mathcal{L}_{0}\left [ \chi  \right ]
\end{align}
where \(\mathcal{L}_{0}\) denotes the linearized collision operator
\begin{equation}\label{L0}
\begin{split}
    \mathcal{L}_{0} [\chi] \equiv &
    -E_p^{-1} \int dP'dP_{1}dP_{2}  \, f_{0}(x,p') \, W_{p,p'\to p_{1},p_{2}} \\
    & \times \bigl( \chi(x,p_{1}) + \chi(x,p_{2}) 
      - \chi(x,p) - \chi(x,p') \bigr),
\end{split}
\end{equation}
with $E_p\equiv u\cdot p$. Note it can be verified that $\mathcal{L}_{0}$ is self-adjoint and positive semidefinite \cite{DeGroot:1980dk} when interpreted as a linear operator in square-integrable Hilbert space. Furthermore, $\mathcal{L}_{0}$ respects the collision invariance as the full  collision kernel does; i.e., $\mathcal{L}_{0}$ has fivefold degenerate zero eigenvalues,
 \begin{align}
 \label{zero}
 \mathcal{L}_0[\psi]=0,   \quad \psi=a+b\cdot p,
 \end{align}
 where $a$ and $b^\mu$ are the free parameters independent of $p^\mu$.

Simultaneously, by projecting  the four-momentum \(p^{\mu}\) onto the temporal and spatial directions (aligned with and orthogonal to the fluid velocity $u^\mu$), the left-hand side (lhs) of the linearized Boltzmann equation  can be conveniently rewritten as
\begin{align}\label{LHS}
    \text{lhs}=E_{p}\left(Df\left ( x,p \right ) +E_{p}^{-1}p^{\left \langle \nu \right \rangle }\partial _{\nu }
f\left ( x,p \right ) \right)
\end{align}
where \(D \equiv u \cdot \partial\) and  \(p^{\left\langle \nu \right\rangle} \equiv \Delta^{\nu \rho} p_{\rho}\). From Eqs.\eqref{rhs} and \eqref{LHS}, we obtain 
\begin{align}\label{pmu}
   E_p Df\left ( x,p \right ) +p^{\left \langle \nu \right \rangle }\partial _{\nu }
f\left ( x,p \right )=-E_pf_{0}\left ( x,p \right ) \mathcal{L}_{0}\left [ \chi  \right ]
\end{align}
where $f$ is regarded as  \(f = f_{0}(1 + \chi)\). 

When it comes to the normal mode analysis, it should be noted that the local equilibrium distribution function is generally not a particular solution of the nonlinear Boltzmann equation, i.e., Eq.(\ref{eq:Boltzmann}). Rather, it merely causes the collision term to vanish. Therefore, the expansion mentioned above cannot be regarded as a perturbation around an actual solution of the equation. A local equilibrium distribution becomes a particular solution of the Boltzmann equation only when it fulfills additional constraints corresponding to global equilibrium $f_{\text{eq}}$, which requires the left-hand side of the equation to also vanish, i.e., \(p^{\mu}\partial_{\mu}f_{0}=0\) resulting in the following Killing conditions:
\begin{align}
\partial_{\left(\mu\right.}\beta_{\left.\nu\right)}=0,\ \ \ \ \ 
\alpha=\text{const}.
\end{align}

In the context of normal mode analysis, we are effectively studying the system's linear response to an external perturbation. Specifically, we consider a small deviation from a given solution of the Boltzmann equation and analyze the subsequent dynamical evolution in response to this initial disturbance, from which one can probe the physical information encoded in the given state described by this particular solution. This also implies that the linearized Boltzmann equation derived from an expansion around the local equilibrium distribution must be appropriately adjusted to proceed with the normal mode analysis. It is well established that the global equilibrium state not only constitutes a particular solution of the Boltzmann equation, but also represents a dynamically stable fixed point \textcolor{red}{\cite{DeGroot:1980dk}}. Consequently, performing a normal mode analysis around this state provides a meaningful way to uncover its underlying physical properties. Therefore, we choose to expand the distribution function around the global equilibrium state as \(f=f_{\text{eq}}(1+\chi)\). Consequently, Eq.\eqref{pmu} can be reformulated as 
\begin{align}\label{partial}
   p^0 \partial_{t}\chi \left ( x,p \right )   +
\mathbf{p}  \cdot\nabla\chi\left ( x,p \right ) 
=-p^0\mathcal{L}_{0}\left [ \chi  \right ],
\end{align}
where we confine our discussion to the rest configuration defined by \(u^\mu=(1,0,0,0)\) in the subsequent calculations. Here, \(u \cdot \partial = \partial_t\), and \(\mathbf{p}\) denotes the spatial component of $p^\mu$. Note that in Eq.\eqref{partial}, \(f_{0}\) should be replaced with \(f_{\text{eq}}\) accordingly in the definition of $\mathcal{L}_0$. Though the analysis performed in this work is restricted to rest configuration, one can naturally extend it to a moving background consistent with the Killing conditions.

\section{A  NOVEL RELAXATION TIME MODEL}
\label{model}

 Although Eq.(\ref{partial}) is a linearized equation, it remains a complicated integro-differential equation, which typically requires numerical methods for its solution;  see, e.g., Refs.~\cite{Moore:2018mma,Ochsenfeld:2023wxz}. In very limited cases, such problems can be treated analytically or semianalytically \cite{Rocha:2024cge}. The difficulty primarily arises from the structure of the collision kernel. A viable approximation to the collision integral would greatly simplify the analysis, allowing for a more general physical discussion. To that end,  we temporarily neglect spatial dependence of the perturbations; thus, we have
  \begin{align}
  \partial_{t}\chi \left ( t,p \right )   
  =-\mathcal{L}_{0}\left [ \chi \left ( t,p \right )    \right ].
  \end{align}
  The operator \(\mathcal{L}_{0}\) does not explicitly contain \(t\); thus,  the formal solution can be obtained
  \begin{align}
  \chi \left (t,p \right ) =e^{-\mathcal{L}_{0}t } \left[\chi \left ( 0,p \right )\right]\label{eq:Liouville}.
  \end{align}
 
 As illustrated in Eq.\eqref{eq:Liouville}, the initial perturbation exhibits an exponential decay in time as a result of  particle collisions, with the decay rate determined by the spectrum of  \(\mathcal{L}_{0}\). If the perturbation is expanded in terms of the eigenfunctions of \(\mathcal{L}_0\), modes with smaller eigenvalues will persist for longer times. Consequently, these modes play a more significant role in the long-time evolution of the system. In particular, these zero-eigenvalue modes correspond to the conserved charges of the system and directly determine its hydrodynamic evolution. Given Eq.(\ref{zero}), the zero modes correspond to the collision invariants in the system, i.e.,
 \begin{align}
 \mathcal{L}_{0}\left | \lambda_{n}   \right \rangle =
 0\ \ \ \ \ \ \ \ \ 
 \left (  n=1,2,3,4,5 \right ) 
 \end{align}
 where $\left | \lambda_{n}   \right \rangle $ denotes the eigenvector of $\mathcal{L}_{0}$. The specific form of the first five eigenvectors can be written as
 \begin{align}
 \left | \lambda _{1}   \right \rangle=1,\ \ \ \ \ 
 \left | \lambda _{2}   \right \rangle=u\cdot p,\ \ \ \ \ 
 \left | \lambda_{3,4,5}   \right \rangle=l \cdot p,\  j \cdot p,\   s \cdot p,
 \end{align}
  where, \(u\), \(l\), \(j\), and \(s\) are pairwise orthogonal, ensuring that \(l^2 = j^2 = s^2 = -1\). In the rest frame, the triad  $(l^\mu,j^\mu,s^\mu)$ represents unit vectors aligned with the $z, x, y$ directions, respectively, such that
 \begin{align}
 \left | \lambda_{1}   \right \rangle=1\ \ \ \ \ 
 \left | \lambda_{2}   \right \rangle=p_0\ \ \ \ \ 
 \left | \lambda _{3,4,5}   \right \rangle=p_{z},\  p_{x},\   p_{y}.
 \end{align}
 
 As mentioned above,  these zero modes  must be retained in any subsequent analysis. Additionally, we also keep the smallest nonzero eigenvalue, which retains the minimal amount of dynamical information necessary for capturing the essential timedependent behavior of the system. In other words, the smallest nonzero eigenvalue is used as a representative of all nonzero eigenvalues, i.e., approximating all nonzero eigenvalues by the smallest one, which can be viewed as replacing the various relaxation timescales with a single, longest relaxation time.  
 The approximated collision operator is now written as
\begin{align}
\label{rta}
    -\mathcal{L}_{0}\simeq -\mathcal{L} = -\left ( 0 \times\sum_{i = 1}^{5}
\left | \lambda _{n}  \right \rangle \left \langle \lambda _{n}  \right | 
+\gamma _{s}  \sum_{i>5}^{}
\left | \lambda _{n}  \right \rangle \left \langle \lambda _{n}  \right | \right )
= -\gamma _{s}+\gamma _{s} \sum_{i = 1}^{5}
\left | \lambda _{n}  \right \rangle \left \langle \lambda _{n}  \right |,
\end{align}
where $\gamma_s$ represents the smallest nonzero eigenvalue. It is easily verified that the justification of this approximation requires a discrete (at least gapped) eigenvalue spectrum of $\mathcal{L}_0$ \cite{Hu:2024tnn}. A keen reader will quickly recognize that the first term on the rhs of Eq.(\ref{rta}) is nothing but the well known relaxation time approximation (by identifying $\gamma_s$ with $1/\tau_R$), also known as the relativistic BGK or Anderson-Witting RTA model \cite{1974Phy....74..466A}. The second term, in the form of a counterterm, is introduced to maintain the crucial property of collision invariance lacking in the traditional RTA, because one can immediately check that 
 \begin{align}
 \mathcal{L}\left | \lambda_{n}   \right \rangle =
 0,\quad \quad \left (  n=1,2,3,4,5 \right ) .
 \end{align}
 This form of approximation has appeared in previous literature referred to as the novel RTA \cite{Rocha:2021zcw,Hu:2022xjn,Hu:2022mvl}. Our construction presented above provides  a new derivation based on eigenvalue spectrum truncation, see also \cite{Hu:2024tnn}. 


    Adopting the novel RTA, and  performing the Fourier transformation to \eqref{partial} would lead to our  main equation for analysis. Considering a plane wave perturbation 
    \(\chi \sim \tilde{\chi}  \left ( k,p \right ) e^{-ik\cdot x} \), 
    the spatial and temporal derivative operators can be replaced by \(\nabla \to i\mathbf{k} \) and
     \(\partial_{t} \to -ik^0\), and a dimensionless equation can be readily obtained
    \begin{align}
        \tau \omega\tilde{\chi}+\hat{p}^{\mu} \kappa _{\mu }   \tilde{\chi}  =-iL\tilde{\chi}\label{eq:Fourier001}
    \end{align}
    with $L[\tilde{\chi } ]\equiv \frac{\tau}{T} \mathcal{L}[\tilde{\chi }]$, or equivalently,
    \begin{align}
    \label{L}
        L[\tilde{\chi } ]\equiv 
\gamma \tau \left (\tilde{\chi }-\sum_{n=1}^{5}\left ( \tilde{\psi }_{n} ,
\tau \tilde{\chi }   \right )  \tilde{\psi }_{n} \right ),
    \end{align}
    where \(\gamma\) is the inverse of the relaxation time scaled by the temperature, namely, 
    \(\gamma \equiv \frac{\gamma_{s} }{T} \), and we introduce two dimensionless shorthand notations for the particle momentum, i.e., $\tau \equiv \frac{p\cdot u}{T} , \hat{p}^{\mu }   \equiv \frac{p^{\mu }}{T}$. Also we decompose the dimensionless Fourier momentum into temporal and spatial components with  \(\Delta^{\mu\nu} \equiv g^{\mu\nu} - u^\mu u^\nu\), shown as
    \begin{align}
      \omega \equiv \frac{k\cdot u}{T} ,\quad 
\kappa^{\alpha }  \equiv \frac{\Delta ^{\alpha \beta }k_{\beta }  }{T} ,
\quad \text{with}\quad \kappa \equiv \sqrt{-\kappa\cdot \kappa}.
    \end{align}
    Without loss of generality, we align the orientation of $\kappa$ with the $z$ direction, i.e., $l^{\alpha } \equiv\frac{\kappa^\alpha }{\kappa}$.  

    In Eq.(\ref{L}), the double bracket representation  is defined as
    \begin{align}
        \left ( B,C \right ) =\int dP\mathit{f_{\text{eq}} } \left ( p \right ) B\left ( p \right ) C\left ( p \right ) .
    \end{align}
    Using the definition of the inner product, we apply the Schmidt orthogonalization procedure to determine the orthogonal and normalized functions $\tilde{\psi}_n$ appearing in Eq.(\ref{L}).  Formally, the expression for \(\tilde{\psi}_i\) takes the same form as that for \(\left| \lambda_i \right\rangle\), but satisfies the orthogonality and normalization conditions \(\left(\tilde{\psi}_i, \tau \tilde{\psi}_j\right) =  \delta_{ij}\). 
    
    The resulting set of  orthogonal eigenfunctions can be constructed with the given \(\left| \lambda_i \right\rangle\)
     \begin{align}
     \tilde{\varphi}_1=1, \quad \tilde{\varphi}_2=\beta (u\cdot p-\frac{e}{n}), \quad \tilde{\varphi}_3=\beta l\cdot p, \quad \tilde{\varphi}_4=\beta j\cdot p, \quad \tilde{\varphi}_5=\beta s\cdot p,
     \end{align}
     satisfying \(\left(\tilde{\varphi}_i, \tau \tilde{\varphi}_j\right) \sim \delta_{ij}\), while \(e\) and \(n\) represent the energy density and particle number density
\begin{align}
    e=\frac{T^{4}z^{4}}{2\pi^{2}}(\frac{1}{z}K_{3}-\frac{1}{z^{2}}K_{2})\exp (\frac{\mu}{T}),\ \ \ \ 
    n=\frac{T^{3}z^{2}}{2\pi^{2}}\exp(\frac{\mu}{T})K_{2} ,
\end{align}
where \(K_{n}\left (z \right )\) is the modified Bessel function of the second kind, defined as follows
\begin{align}
    K_{n} \left ( z \right ) \equiv \int_{0}^{\infty } dx\ \cosh \left ( nx \right ) 
e^{-z\cosh x} .
\end{align}
When nothing confusing occurs, the dependence on $z$ in $K_n(z)$ will be suppressed below for compactness.

     Furthermore, the orthogonal and normalized eigenfunctions are 
    \begin{align}\label{psi}
        \tilde{\psi } _{1} =\frac{\tilde{\varphi}_{1}   }{\sqrt{V_{11} } } ,\ \ \ 
\tilde{\psi } _{2} =\frac{\tilde{\varphi}_{2}   }{\sqrt{V_{22} } } ,\ \ \ 
\tilde{\psi } _{3} =\frac{\tilde{\varphi}_{3}   }{\sqrt{V_{33} } } ,\ \ \ 
\tilde{\psi } _{4} =\frac{\tilde{\varphi}_{4}   }{\sqrt{V_{44} } } ,\ \ \ 
\tilde{\psi } _{5} =\frac{\tilde{\varphi}_{5}   }{\sqrt{V_{55} } } ,
    \end{align}
    with the normalization factor $V_{ij}\equiv \int dPf_{\text{eq}} \left ( p \right ) \tilde{\varphi } _{i} 
        \tau \tilde{\varphi } _{j}$.
    For more technical details of Schmidt orthogonalization, one could refer to \cite{Rocha:2021zcw, Hu:2023elg}. With all relevant components in place, one can readily verify that $L[\tilde{\psi}_i]=0$ for $i=1,2,3,4,5$, and thus the collision invariance is recovered as expected.

\section{Normal mode analysis}
\label{main}

In this section, we give a detailed normal mode analysis of kinetic theory in the relaxation time approximation. Our discussion follows from a similar study \cite{Hu:2023elg}; however, we relax the ultrarelativistic assumption therein to account for massive transport. A more complete approach  would involve calculating the system's retarded correlation functions \cite{Romatschke:2015gic,Katz:2019rgf}; see recent calculations on massive transport \cite{Hataei:2025mqf,Bajec:2025dqm}. The analysis we present here only recovers the information contained in the denominator of the retarded correlators. However, this is already sufficient for the purposes of normal mode analysis. At the end of  this section, 
we will give several brief comments and compare the two mentioned approaches.

    To facilitate the following discussion, we define the fluctuation amplitudes as
    \begin{align}
        \rho _{n} \left ( k \right ) \equiv \left ( \tilde{\psi }_{n} ,\tau \tilde{\chi} \left ( k,p \right ) \right ) ,\label{eq:rhon}
    \end{align}
   where $n$ runs from $1$ to $5$ with an exception for \(\rho_{2}\),
    \begin{align}
        \rho _{2} \left ( k \right ) \equiv \frac{1}{\sqrt{V_{22} } } \left ( \tau ,\tau \tilde{\chi} \left ( k,p \right ) \right ).  \label{eq:rho2}
    \end{align}
Note that $ \tilde{\chi} \left (k,p \right )$ represents the deviation of the distribution from equilibrium, and accordingly, $\rho_n$ can be naturally interpreted as the fluctuation amplitudes of the conserved charge densities. This is precisely why $\rho_2$ is defined in the given form — it represents a well defined fluctuation of the energy density, differing from it at most by a constant factor.

    Once  \(\rho _{n} \left ( k \right )\) is given, we can  express Eq.\eqref{eq:Fourier001} as
    \begin{align}\label{chi}
        \left ( c+\frac{\hat{p} }{\tau }\cdot l  \right )\tilde{\chi }=
        \hat{\gamma }\sum_{n=1}^{5}\left ( \tilde{\psi }_{n},\tau\tilde{\chi }  \right ) \tilde{\psi }_{n}=
        \hat{\gamma }\sum_{n=1}^{5}\rho _{n}\bar{\psi }_{n},
    \end{align}
where  Eq.\eqref{eq:rho2} is utilized to rewrite the rhs of Eq.\eqref{eq:Fourier001}, and we have introduced the abbreviation
    \begin{align}
    \label{main1}
    c\equiv \frac{\omega }{\kappa}+ i\frac{\gamma}{\kappa}  , \ \ \ \hat{\gamma}\equiv i\frac{\gamma }{\kappa}.
    \end{align}
 Note we introduce a new set of basis $\{\bar{\psi}_i\}$ defined as
    \(\bar{\psi } _{1} =\tilde{\psi } _{1} -\frac{\beta e\sqrt{V_{11} } }{nV_{22} }\tau 
+ \frac{\beta^{2}  e^{2} \sqrt{V_{11} } }{n^{2} V_{22} }\) and \(\bar{\psi } _{i}=\tilde{\psi } _{i}\) for \(i\neq1\), to coordinate the definition of $\rho_2$.
    Then, $\tilde{\chi}$ can be determined from Eq.(\ref{main1})
    \begin{align}
        \tilde{\chi } =\frac{\hat{\gamma}   {\textstyle \sum_{i=1}^{5}}   \rho _{i} \bar{\psi} _{i} }
        {c+\frac{\hat{p}  }{\tau  } \cdot l} \label{eq:rho&kai}.
    \end{align}
Note that the inversion can be safely made only when \(\left(c + \frac{\hat{p}}{\tau} \cdot l \right) \neq 0\). If not, we cannot reach Eq.(\ref{eq:rho&kai}). As a result, the discussion splits into two cases depending on whether this equation  \(\left(c + \frac{\hat{p}}{\tau} \cdot l \right) = 0\)  holds as will be discussed later.
    
    \subsection{Collective modes}\label{collective}

First, we discuss the case of  \(\left(c + \frac{\hat{p}}{\tau} \cdot l \right) \neq 0\). 
By substituting Eq.\eqref{eq:rho&kai} into Eqs.\eqref{eq:rhon} and \eqref{eq:rho2}, we obtain

    \begin{align}\label{eq:rho}
        \rho _{i}=\sum_{j=1}^{5} \int dP f_{\text{eq}}  \left ( p \right ) 
        \frac{\tilde{\psi}  _{i}\hat{\gamma}  \tau \rho _{j} \bar{\psi} _{j} }
        {c+\frac{\left |  \mathbf{p} \right | }{p^{0} } \cos\theta } ,
    \end{align}
    where $\cos\theta\equiv \frac{p_z}{| \mathbf{p}|}$ and $i$ runs from $1$ to $5$. Equation (\ref{eq:rho}) can be casted into a more concise form
    \begin{align}\label{Arho}
    \sum_{j=1}^{5}A_{ij}\rho_j=0,
    \end{align}
    by introducing \(A_{ij}\) to denote the coefficients of the secular equation in terms of \(\rho_{n}\)
    \begin{align}\label{Aij}
        A_{ij}=\int dP f_{\text{eq}}\left ( p \right )\frac{\tilde{\psi }_{i}\hat{\gamma} \tau \bar{\psi }_{j}   }
    {c+\frac{\left |  \mathbf{p} \right | }{p^{0} } \cos\theta }    
    -\delta _{ij} .
    \end{align}

    The vanishing determinant of the coefficient matrix \(A_{ij}\) in the secular equation corresponds to the solvability condition for the system of linear homogeneous equations. According to this principle, $\det A_{ij}=0$ gives rise to three block-diagonal sectors
    \begin{align}
       \Phi_{1}\left ( \hat{\gamma },c,z  \right ) =\det
        \begin{bmatrix}
            A_{11} & A_{12} & A_{13}\\
            A_{21} & A_{22} & A_{23}\\
            A_{31} & A_{32} & A_{33}
        \end{bmatrix} \label{eq:Phi_1}
        =0\ \ ,\ \ 
        \Phi_{2}\left ( \hat{\gamma },c,z  \right ) =
        A_{44}=A_{55}
        =0,
    \end{align}
    where $z$ denotes the scaled mass $m/T$. As is shown clearly, the sectors for \(\rho_4\) and \(\rho_5\) degenerate, and decouple from the sector for determining \(\rho_1\), \(\rho_2\), and \(\rho_3\). We refer to the $\Phi_1$ sector as the one in which the sound and heat channels are coupled, and the $\Phi_2$ sector as the shear channel. In the long-wavelength limit, the corresponding secular equations will yield the hydrodynamic modes, as we shall demonstrate in the following. Compared to the results in \cite{Romatschke:2015gic,Hu:2023elg},  the sound and heat channels are coupled to each other, exhibiting sound-heat coupling. Generally, in a transport system with nonzero chemical potential, the sound and heat (or charge) channels are coupled. The decoupling of these two channels reported in Refs.\cite{Romatschke:2015gic,Hu:2023elg} in the massless limit reflects the simplification inherent in the RTA. It is an artifact of the approximation, not a genuine physical phenomenon. If the model is slightly modified, for instance, by introducing an energy-dependent relaxation time, the decoupled channels would once again couple as expected in the presence of a nonzero chemical potential. 
   
    Since the integration of \(A_{ij}\) in spherical coordinates involves a triple integral over momentum magnitude, azimuthal angle, and polar angle, it is quite challenging. However, by reordering the integrals, performing the angular integrations first and leaving the momentum magnitude for last, the complexity can be effectively condensed into the expression involving the momentum integral. This leads us to a  common structure arising in $A_{ij}$,
\begin{align}\label{eq:G}
    G\left ( c,z,m,n \right ) = \int_{0}^{\infty } \exp\left ( -\sqrt{z^{2}+\mathbf{\left | \hat{p} \right | }^{2}   }  \right ) 
    \mathbf{\left | \hat{p} \right | }^{n} 
    \left ( z^{2}+\mathbf{\left | \hat{p} \right | }^{2}  \right ) ^{\frac{m}{2} } 
    \coth^{-1} \left ( c\frac{\sqrt{z^{2}+\mathbf{\left | \hat{p} \right | }^{2}   }}{\left | \mathbf{\hat{p}}  \right |}  \right ) 
    d\left | \mathbf{\hat{p}}  \right | .
\end{align}
The value of two integers \(m\) and \(n\) depends on the specific form of \(A_{ij}\). \(A_{ij}\) can be expressed as a linear combination of the function \(G\) and an analytical expression independent of $G$. For example, for \(A_{13}\),
\begin{align}
    A_{13}=\hat{\gamma}  \left ( \frac{z^{2} \left ( zK_{3} -K_{2} \right ) }{\sqrt{z^{5} K_{2} K_{3} } }  
-\frac{cG\left ( c,z,2,1 \right )}{\sqrt{z^{5} K_{2} K_{3} }} 
 \right ).
\end{align}

The complexity in addressing the secular equation \eqref{eq:Phi_1} stems from the presence of four free parameters $\hat{\gamma}, z$, and complex-valued $c$.  Nevertheless, our present aim is not to compute the fluctuation amplitudes $\rho_n$, but instead to derive the conditions for the existence of collective modes at intermediate and large wave numbers. As is exhibited in  \cite{Romatschke:2015gic,Hu:2023elg}, there is a sharp transition behavior from a region supporting collective modes to one where such modes are absent at some critical value $\kappa_c$. When the above discussion is extended to massive systems, one may naturally ask whether the aforementioned transition behavior still persists. If so, do the characteristics  of this transition change? And does it exhibit any nontrivial dependence on the mass parameter? These are the key questions we aim to investigate in the following analysis.

According to the argument principle in complex analysis (see Appendix \ref{argu}), the number of zeros of \(\Phi\) within a region of the complex $c$ plane in which $\Phi$ is analytic is equal to the number of times the image of \(\Phi\),  evaluated along a closed contour enclosing that region, winds around the origin in the complex plane. For a given mass \(z\), we fix \(\hat{\gamma}\) and vary \(c\) to trace the trajectory of \(\Phi\). To determine the number of zeros of $\Phi$ in the complex plane, we construct a closed contour surrounding the relevant region in the $c$ plane. This contour consists of two parts: one that closely follows the real axis from above (without intersecting it) and another that forms a large semicircular arc in the upper half plane, effectively enclosing the entire upper half plane. Similarly, a separate contour can be defined in the lower half plane. Note that the nonanalytical structures of $\Phi$ can only appear when \(c\) is a real number and we have checked that there are no zeros in this case.

\begin{figure}[H]
    \centering
    \includegraphics[scale=0.7]{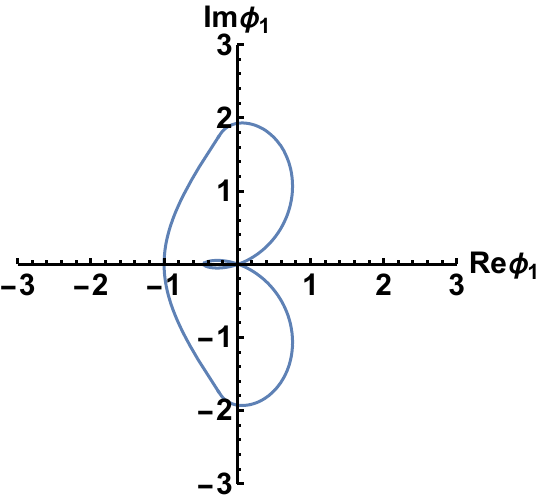}\ \ \ \ \ \ \ \ \ \ \ \ \ \ 
    \includegraphics[scale=0.7]{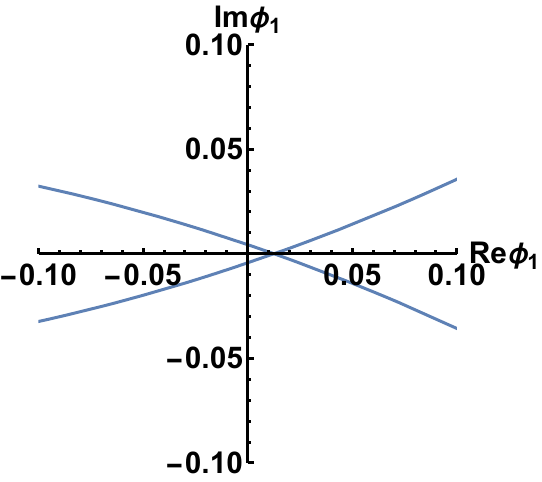}
    \caption{The typical trajectory of \(\Phi_{1}\left(0.22i,c,1\right)\). The right panel is an enlarged view of the left panel near the origin. The curve starts at \((-1,0)\), as \(c\) moves from \(-\infty +i 0^{+}\) to \(\infty +i 0^{+}\) and returns to \((-1,0)\) after encircling the origin twice.}
    \label{fig:traject1}
\end{figure}

\begin{figure}[H]
    \centering
    \includegraphics[scale=0.7]{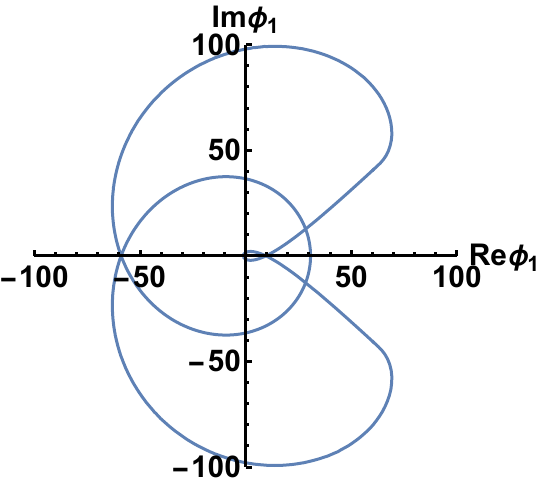}\ \ \ \ \ \ \ \ \ \ \ \ \ \ 
    \includegraphics[scale=0.7]{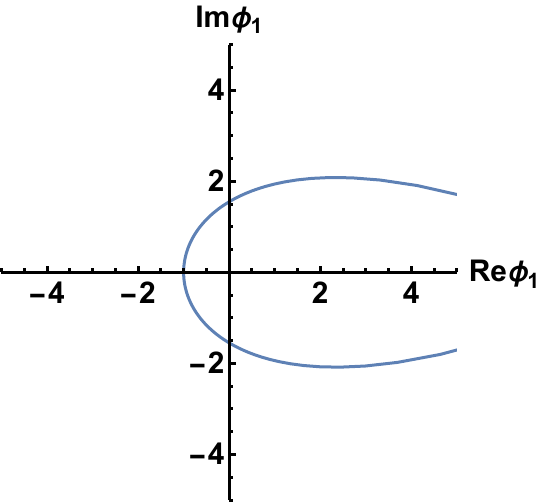}
    \caption{The typical trajectory of \(\Phi_{1}\left(2i,c,1\right)\). The right panel is an enlarged view of the left panel near the origin. The curve starts at \((-1,0)\), as \(c\) moves from \(-\infty +i 0^{+}\) to \(\infty +i 0^{+}\), and returns to \((-1,0)\) after encircling the origin thrice.}
    \label{fig:traject2}
\end{figure}

First, we focus on the asymptotic behavior of \(\Phi\) as \(c\to \infty\), which is easily obtained,
\begin{align}
    \Phi_{1}\left(\hat{\gamma},c\to \infty,z\right)=\Phi_{2}\left(\hat{\gamma},c\to \infty,z\right)=-1.
\end{align}
This significantly simplifies the analysis of the trajectory. As a result, we only need to focus on the motion of $c$ from \(-\infty +i 0^{+}\) to \(\infty +i 0^{+}\), and determine how many times it encircles the origin. As illustrated in Figs. \ref{fig:traject1} and \ref{fig:traject2}, the number of zeros varies with $\hat{\gamma}$.



For a fixed mass, we observe that the number of zeros increases with   \(\hat{\gamma}\) because  \(\hat{\gamma}\) is inversely proportional to the wave number. As \(\hat{\gamma}\) is varied numerically, the trajectory in the complex plane evolves gradually until it encircles the origin. The corresponding value of \(\hat{\gamma}\) at this point is referred to as the critical value. In the case of \(\Phi_{1}\), we identify two distinct critical values, each corresponding to a separate transition in the winding number around the origin. These transitions are associated with the onsets of the heat and sound channels, respectively. As shown in Fig.\ref{fig:traject1}, when \(c\) moves along the curve in the upper half plane, the trajectory of \(\Phi_{1}\) encircles the origin twice, corresponding to two collective modes (or hydrodynamic modes): the two sound modes. From more detailed subsequent critical value analysis, these two sound modes appear simultaneously. Subsequently, as shown in Fig.\ref{fig:traject2}, the trajectory of \(\Phi_{1}\) encircles the origin three times, indicating the emergence of another collective mode: the heat mode. Furthermore, an analysis of \(\Phi_{2}\) reveals the emergence of a single mode—identified as the shear mode—once  \(\hat{\gamma}\) surpasses the critical threshold. Notably, when the integration contour is chosen in the lower half plane of the complex $c$ plane, no zeros are observed throughout the entire trajectory.

\begin{figure}[H]
    \centering
    \includegraphics[scale=0.26]{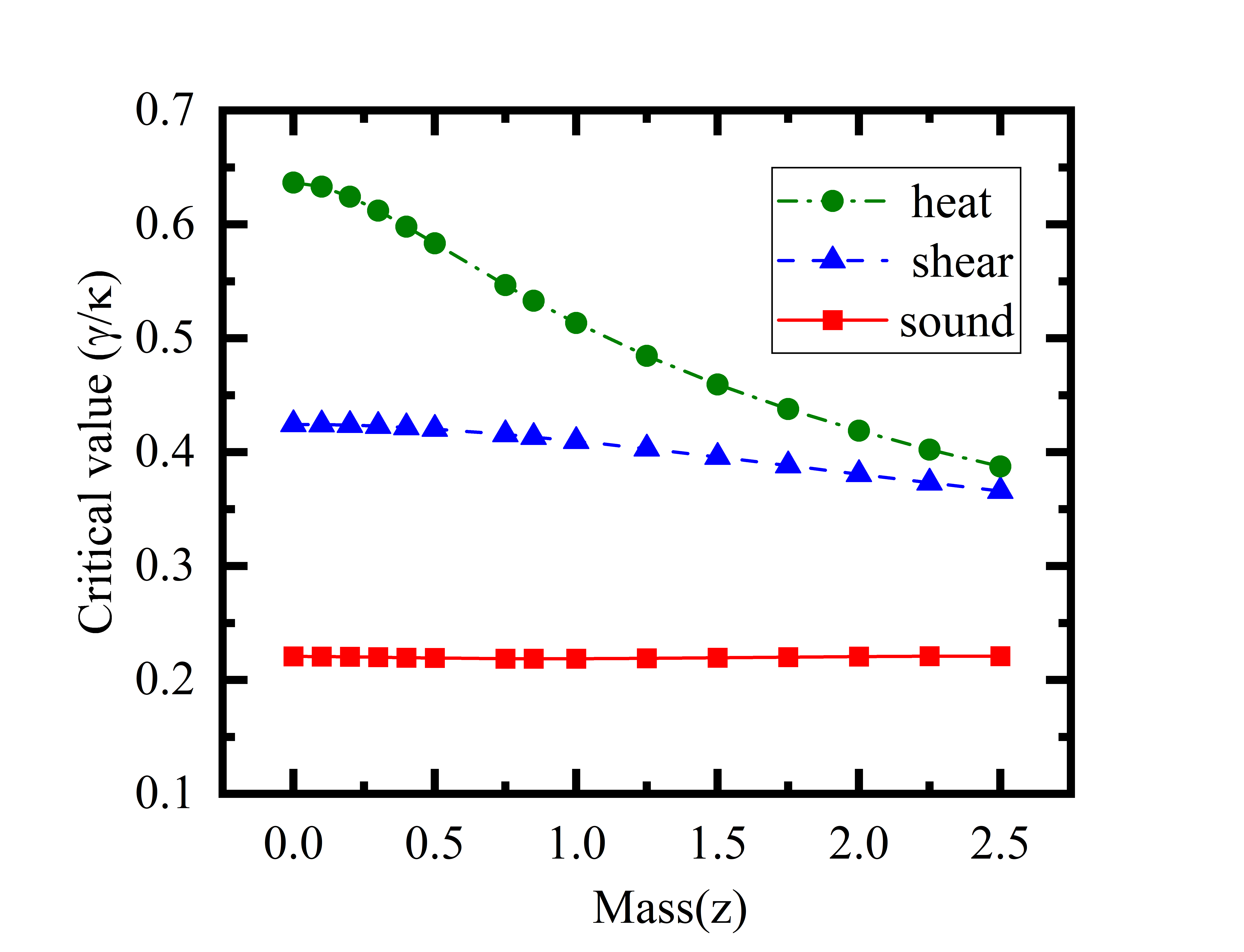}\ \ \ \ 
    \includegraphics[scale=0.26]{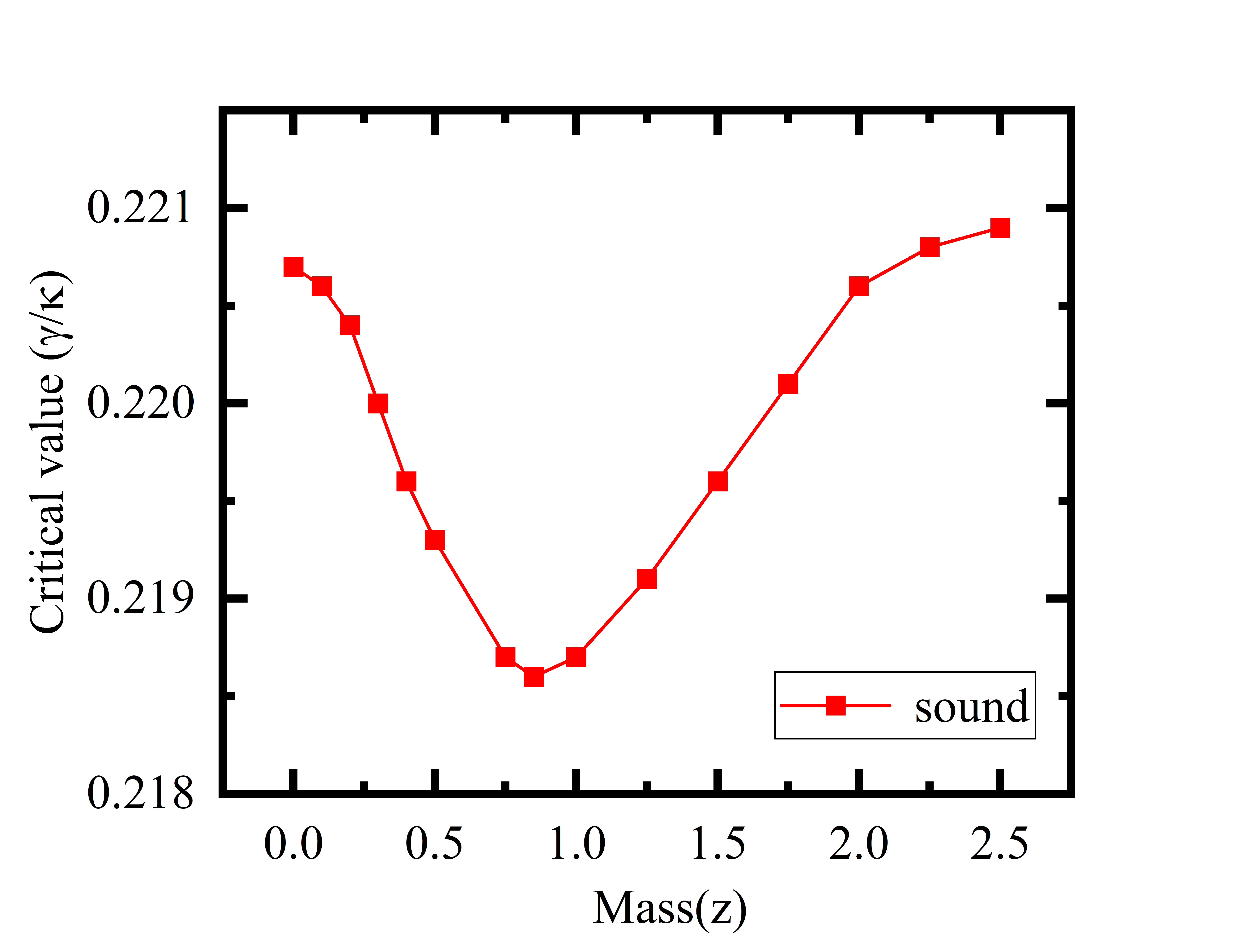}
    \caption{The critical values of \(\frac{\gamma}{\kappa}\) for various collective modes. Left panel: the critical values of \(\frac{\gamma}{\kappa}\) for all collective modes. Right panel: enlargement of the sound mode behavior. All values are determined numerically to precision \(10^{-4}\).}
    \label{fig:modes}
\end{figure}

 Our results indicate that the number of zeros depends sensitively on the value of  \(\hat{\gamma}\), and we can determine the critical thresholds at which new solutions appear accordingly. In Fig.\ref{fig:modes}, the critical wave numbers for these collective modes corresponding to different scaled masses are given.  In particular, for \(z=0\), our results are consistent with those reported in \cite{Hu:2023elg}. Within the range of scaled masses considered, the critical wave numbers for the heat and shear channels increase with increasing scaled mass. One can find a reasonable explanation for this trend: The critical wave number $\kappa_c$ signifies the threshold above which hydrodynamic collective modes, such as sound waves, are no longer sustained. 
 Once these collective modes are formed, it signifies that the constituent particles of the system have achieved organized coordinated motion on a macroscopic scale. The wave number $\kappa$ can serve as a measure of the perturbation strength imposed by the system's free-stream dynamics on the collective motion. When the wave number is large, this corresponds to a short-wavelength perturbation, effectively disturbing only localized subsets of particles and causing them to deviate from the collective motion. If the particles have a large mass (inertia), they will remain largely unaffected and maintain the collective motion, implying that a perturbation with a larger $\kappa$ is required to disrupt it. Conversely, if the particle mass is small, the particles are more susceptible to short-wavelength perturbations. This provides a qualitative explanation for the observed dependence of collective modes in the shear channel and heat conduction channel on the (scaled) particle mass. However, the sound channel exhibits a nonmonotonic dependence on the scaled mass; see the right panel of Fig.\ref{fig:modes}. 
Although we have not found a satisfying reason for it, the behavior of the sound channel exhibits a less distinct trend compared to the shear and heat channels (see the left panel).
Furthermore, the figure reveals a clear hierarchy in the stability of the modes against inhomogeneous perturbations: The sound mode exhibits the greatest resilience to perturbations, remaining well defined even at large $\kappa$; the shear mode follows, and the heat mode is the least stable, disappearing at relatively smaller wave numbers. 
 


   In the preceding analysis, we have established the conditions for the existence of solutions and determined the critical wave numbers below which the collective modes survive. As is shown clearly, $\kappa \leq \kappa_c <1$, these collective modes are hydrodynamic in nature. We now proceed to analytically derive the collective modes and obtain the corresponding dispersion relations in the long-wavelength limit.
   We focus on the regime  \(\omega\propto \kappa\ll 1\), in which we set \(\omega=b \kappa\) with \(b\) a complex parameter of order $O(1)$. By expanding in powers of \(\kappa\) and substituting back for \(\omega\) , the integral involving  \(\coth^{-1} x\) in Eq.\eqref{eq:G} reduces to a polynomial form, allowing for an analytical treatment while preserving the leading-order terms in \(\kappa\). 


Then, we can proceed to handle the integral with respect to the momentum magnitude. First, we provide the specific form of each element in the determinant represented by \(\Phi_{1}\left ( \kappa,\omega,z\right )\) in Eq.\eqref{eq:Phi_1}: 
    \begin{equation}\label{eq:determinant}
\begin{aligned}
A_{11} &= \frac{-3z^{2}(\kappa^{2}+\omega(\omega-i\gamma))K_{1}^{2}
          +K_{2}(Ki_{1}\kappa^{2}z(12+z^{2})
          +(3+z^{2})(\kappa^{2}+3\omega(\omega-i\gamma))K_{2})}
         {3\gamma^{2}(z^{2}K_{1}^{2}+3zK_{1}K_{2}-(3+z^{2})K_{2}^{2})} \\
       &\quad -\frac{z K_{1}(-3Ki_{1}\kappa^{2}z+(\kappa^{2}(12+z^{2})
          +9\omega(\omega-i\gamma))K_{2})}
         {3\gamma^{2}(z^{2}K_{1}^{2}+3zK_{1}K_{2}-(3+z^{2})K_{2}^{2})}, \\
A_{12} &= \frac{\kappa^{2}(-z K_{1}^{2}+K_{1}(zKi_{1}-2K_{2})+3Ki_{1}K_{2})}
         {3\gamma^{2} K_{2}\sqrt{-(z^{2}K_{1}^{2}+3zK_{1}K_{2}-(3+z^{2})K_{2}^{2})}}, \\
A_{13} &= \frac{\kappa(-2\omega+i\gamma)K_{2}}
         {\gamma^{2}\sqrt{K_{2}(zK_{1}+4K_{2})}}, \\
A_{21} &= \frac{\kappa^{2}zK_{2}(zK_{1}^{2}+4K_{1}K_{2}-z K_{2}^{2})}
         {\gamma^{2} \left (\sqrt{-(z^{2}K_{1}^{2}+3zK_{1}K_{2}-(3+z^{2})K_{2}^{2})}\right )^{3}}, \\
A_{22} &= \frac{-\omega z^{2}(\omega-i\gamma) K_{1}^{2}-3\omega z(\omega-i\gamma) K_{1}K_{2}
          +(\kappa^{2}+\omega(3+z^{2})(\omega-i\gamma)) K_{2}^{2}}
         {\gamma^{2} (z^{2}K_{1}^{2}+3zK_{1}K_{2}-(3+z^{2})K_{2}^{2})}, \\
A_{23} &= \frac{i\kappa(2i\omega+\gamma)(zK_{1}+4K_{2})\sqrt{K_{2}}}
         {\gamma^{2}\sqrt{-z^{3}K_{1}^{3}-7z^{2}K_{1}^{2}K_{2}
          +z(-9+z^{2})K_{1}K_{2}^{2}+4(3+z^{2})K_{1}^{3}}}, \\
A_{31} &= \frac{-\kappa z(2\omega-i\gamma)(zK_{1}^{2}+4K_{1}K_{2}-z K_{2}^{2})\sqrt{K_{2}}}
         {\gamma^{2} (z^{2}K_{1}^{2}+3zK_{1}K_{2}-(3+z^{2})K_{2}^{2})}, \\
A_{32} &= \frac{\kappa(-2\omega+i\gamma)K_{2}^{3/2}}
         {\gamma^{2}\sqrt{-z^{3}K_{1}^{3}-7z^{2}K_{1}^{2}K_{2}
          +z(-9+z^{2})K_{1}K_{2}^{2}+4(3+z^{2})K_{1}^{3}}}, \\
A_{33} &= \frac{Ki_{1}\kappa^{2}z^{3}-z(\kappa^{2}(3+z^{2})
          +5\omega(\omega-i\gamma))K_{1}
          +(\kappa^{2}(-12+z^{2})-20\omega(\omega-i\gamma)) K_{2}}
         {5\gamma^{2}(zK_{1}+4K_{2})},
\end{aligned}
\end{equation}
    where we retain the terms up to $O(\kappa^2)$. In the above expressions,  the Bickley function \(Ki_{n}(z)\) appears defined as \cite{nist}
    \begin{align}
        Ki_{n}\left ( z \right ) \equiv 
         \int_{0}^{\infty } e^{-z\cosh x} 
        \frac{1   }{\cosh ^{n}x} dx.
    \end{align}
    Below, the dependence on $z$ of $Ki_n(z)$ will be suppressed.

    Substituting Eq.\eqref{eq:determinant} into Eq.\eqref{eq:Phi_1} results in a sextic equation in \(\kappa\), which is difficult to solve analytically. To obtain the dispersion relation analytically or semianalytically, we retain terms up to fourth order in $\kappa$ in the equation and neglect contributions from higher-order terms.  The resulting equation is as follows
    \begin{align}
    B_{1}\left(\omega,z\right)+B_{2}\left(\omega,z\right)\kappa^{2}+B_{3}\left(\omega,z\right)\kappa^{4}=0,
    \end{align}
    with the coefficients  expressed as
    \begin{equation}\label{eq:B_functions}
\begin{split}
B_{1}(\omega,z) &= -15i\gamma^{3} \bigl( z^{3}K_{1}^{3} + 7z^{2}K_{1}^{2}K_{2} 
                  - z(-9+z^{2}) K_{1}K_{2}^{2} - 4(3+z^{2}) K_{2}^{3} \bigr) \omega^{3}, \\ 
B_{2}(\omega,z) &= \gamma^{2} \Bigl( 3z^{2}(8+z^{2}) K_{1}^{3} 
                  + z^{2}K_{1}^{2} \bigl( -3zKi_{1}(5+z^{2}) 
                  + (108+11z^{2}) K_{2} \bigr) \\
                &\quad - zK_{1}K_{2} \bigl( 2zKi_{1}(60+7z^{2}) 
                  + 3(-22+4z^{2}+z^{4})+ K_{2} \bigr) \\
                &\quad + K_{2} \bigl( zKi_{1}(-240-11z^{2}+3z^{4}) 
                  + (72+28z^{2}+3z^{4}) K_{2} \bigr) \Bigr) \omega^{2} \\
                &\quad + 15i\gamma^{3}K_{2} \bigl( z^{2}K_{1}^{2} + 3zK_{1}K_{2} 
                  - (4+z^{2})K_{2}^{2} \bigr) \omega, \\
B_{3}(\omega,z) &= 5\gamma^{2} (zK_{1}+4K_{2}) \bigl( -z^{2}K_{1}^{2} 
                  + zK_{1}(zKi_{1}-3K_{2}) + K_{2}(4zKi_{1}+K_{2}) \bigr).
\end{split}
\end{equation}

    We solve for \(\kappa\)  as the dispersion relation of the sound modes
    \begin{align}\label{eq:sound}
        \kappa_{sound}&=
        \sqrt{\frac{-15\gamma K_{2}^{2}\left ( z^{2}K_{1}^{2} +3zK_{1}K_{2}
        -\left ( 4+z^{2} \right ) K_{2}^{2}\right ) ^{2}S_{1}\omega^{2}+
        S_{2}\omega^{3}}{15\gamma K_{2}^{3}\left ( z^{2}K_{1}^{2} +3zK_{1}K_{2}
        -\left ( 4+z^{2} \right ) K_{2}^{2}\right ) ^{3}}}\nonumber\\
        &=
        \begin{cases}
            \sqrt{\frac{3\omega^{2}z\left ( 6i\omega+5\gamma  \right ) }{25\gamma } },\ z\gg 1\\
            \sqrt{\frac{3\omega^{2}\left ( 4i\omega+5\gamma  \right ) }{5\gamma } },\ \ z\ll 1
        \end{cases},
    \end{align}
    with its nonrelativistic and ultrarelativistic limits given. Here, \(S_{1},S_{2}\) are functions only dependent on $z$,
   \begin{equation}\label{eq:S_function}
\begin{aligned}
S_{1} &= (zK_{1}+4K_{2})(-z^{2}K_{1}^{2}-3zK_{1}K_{2}+(3+z^{2})K_{2}^{2}), \\
S_{2} &= iS_{1}\big[5z^{6}K_{1}^{6}+z^{5}K_{1}^{5}(-5zKi_{1}+(46-3z^{2})K_{2}) \\
       &\quad + z^{4}K_{1}^{4}K_{2}(3zKi_{1}(-20+z^{2})-5(-24+5z^{2})K_{2}) \\
       &\quad + 2z^{3}K_{1}^{3}K_{2}^{2}(2zKi_{1}(-60+7z^{2})+(28-20z^{2}+3z^{4})K_{2}) \\
       &\quad + z^{2}K_{1}^{2}K_{2}^{3}(-2zKi_{1}(160-43z^{2}+3z^{4})+(52+55z^{2}+17z^{4})K_{2}) \\
       &\quad + zK_{1}K_{2}^{4}(z^{3}Ki_{1}(97-23z^{2})+(288+4z^{2}-33z^{4}-3z^{6})K_{2}) \\
       &\quad + K_{2}^{5}(z^{3}Ki_{1}(36+z^{2}+3z^{4})+(-192-96z^{2}-5z^{4}+3z^{6})K_{2})\big].
\end{aligned}
\end{equation}
    From Eq.\eqref{eq:sound} in the limit where \(z\gg 1\), the speed of sound for a nonrelativistic gas is reproduced,
    \begin{align}
       \omega=\pm c_s\big|_{z \gg 1}\,\kappa-\frac{i}{z\gamma}\kappa^{2},
    \end{align}
    where $c_s|_{z \gg 1}=\sqrt{\frac{5}{3z}}$ is well known historically. However, the damping rate differs from that given in \cite{cercignani2002relativistic}. Note as an aside, the damping rate in the ultrarelativistic limit presented in \cite{cercignani2002relativistic} is not consistent with Eq.(\ref{sound}).
    
    Also, we obtain  \(\kappa\) for the heat mode,
    \begin{align}\label{eq:thermal}
        \kappa_{heat}&=\sqrt{\frac{3i\gamma K_{2}\left ( z^{2}K_{1}^{2}+3zK_{1}K_{2}-
        \left ( 4+z^{2} \right ) K_{2}^{2} \right ) \omega}
    {\left ( zK_{1}+4K_{2} \right ) 
    \left ( z^{2}K_{1}^{2}-K_{2}\left ( 4zKi_{1} +K_{2}\right ) 
    +zK_{1}\left ( -zKi_{1} +3K_{2}\right )  \right ) } } \nonumber\\&=
    \begin{cases} 
        \frac{1+i}{\sqrt{2} } \sqrt{z\omega\gamma  }, \ \ z\gg 1 \\ 
        \frac{1+i}{\sqrt{2} } \sqrt{3\omega\gamma  }, \ z\ll 1 
        \end{cases},
    \end{align}
    where the above results for the heat mode are consistent with the calculation in \cite{cercignani2002relativistic} except the nonrelativistic limit. Both analytically (via \textit{Mathematica}) and numerically, we confirm that a typographical error  appears in the results of \cite{cercignani2002relativistic}. 

    Similarly, for \(\Phi _{2}\) in Eq.\eqref{eq:Phi_1}, we  also obtain the shear modes,
    \begin{align}\label{eq:shear}
        \kappa_{shear}=
        -\sqrt{\frac{15\omega\left ( \omega-i\gamma  \right )\left ( zK_{1}+4K_{2} \right )  }
        {\left(z^{3}Ki_{1}-\left ( 3z+z^{3} \right ) K_{1}+\left ( z^{2}-12 \right )K_{2}\right) } }
        =
        \begin{cases} 
            \frac{1+i}{\sqrt{2} } \sqrt{z\omega\gamma  } ,\ z\gg 1 \\ 
            \frac{1+i}{\sqrt{2} } \sqrt{5\omega\gamma  } , \ z\ll 1 
            \end{cases},
    \end{align}
    which are consistent with the results in \cite{cercignani2002relativistic}.

    In Eqs.\eqref{eq:sound}, \eqref{eq:thermal}, and \eqref{eq:shear}, we have provided the dispersion relations for these collective modes. By inverting the relation to express $\omega$ as a function of $\kappa$, we can obtain a more conventional form of the dispersion relation. In particular, in the ultrarelativistic limit,  several well known results are recovered \cite{Hu:2023elg,Romatschke:2015gic} and shown as
    \begin{align}
    \label{sound}
        \omega&=\pm c_{s}\big|_{z=0}\,\kappa -\frac{2i}{15\gamma}\kappa^{2},\ \ \ \ \ \ \ \ \text{sound\ \ modes,}\\
        \omega&=-\frac{i}{3\gamma}\kappa^{2},\ \ \ \ \ \ \ \ \text{heat\ \ mode,}\\
        \label{shear}
        \omega&=-\frac{i}{5\gamma}\kappa^{2},\ \ \ \ \ \ \ \ \text{shear\ \ modes,}
    \end{align}
    where the speed of sound approaches its conformal limit \(c_s\big|_{z=0}=\frac{1}{\sqrt{3}}\). Additionally, we find that in this case $z \ll 1$, the sound and heat channels in 
    \(\Phi_{1}\) decouple, resulting in
    \begin{align}
        \Phi_{1}\left(\hat{\gamma},c\right) 
        = \left(\hat{\gamma}\coth^{-1}c -1\right)
        \left(3\hat{\gamma}\left(c-\hat{\gamma}\right)
        +\left(3c\hat{\gamma}\left(
            \hat{\gamma}-c
        \right)-\hat{\gamma}\right)\coth^{-1}c+1\right)=0,
    \end{align}
    as expected in \cite{Hu:2023elg}. Based on the above analysis, the number of obtained modes matches the number of zeros derived from the argument principle. These dynamical modes originate from the zero eigenvalues of the collision operator, reflecting the organized collective motion of particles coordinated by collisions. 

    Equations \eqref{eq:sound}, \eqref{eq:thermal}, and \eqref{eq:shear} together with their ultrarelativistic and nonrelativistic limits  constitute one of the main results of this work. Before ending this subsection, we would like to provide a comprehensive comparison (shown below by Table I) of our results with those given in two recent works \cite{Hataei:2025mqf,Bajec:2025dqm}. 
    
    \begin{table}
\begin{tabular}{|c|c|c|c|c|} 
\hline
 & This work  & \cite{Hataei:2025mqf} & \cite{Bajec:2025dqm} & \cite{cercignani2002relativistic} \\ \hline
Sound (z $\ll$ 1) &$\omega=\pm c_{s}\,\kappa -\frac{2i}{15\gamma}\kappa^{2}$  & ... & $\omega=\pm c^{\prime}_{s}\,\kappa -\frac{2i}{15\gamma}\kappa^{2}$ & $\omega= \pm c_{s}\,\kappa -\frac{i}{6\gamma}\kappa^{2}$  \\ \hline
Sound (z $\gg$ 1)&$\omega=\pm c_{s}\,\kappa -\frac{i}{z\gamma}\kappa^{2}$  & ... & $\omega=\pm c^{\prime}_{s}\,\kappa -\frac{i}{z\gamma}\kappa^{2}$ & $\omega= \pm c_{s}\,\kappa -\frac{i}{z\gamma}\kappa^{2}$ \\ \hline
Heat (z $\ll$ 1)&$\omega=-\frac{i}{3\gamma}\kappa^{2}$ & $\omega=-\frac{i}{3\gamma}\kappa^{2}$ & $\omega=-\frac{i}{3\gamma}\kappa^{2}$ & $\omega=-\frac{i}{3\gamma}\kappa^{2}$ \\ \hline
Heat (z $\gg$ 1)&$\omega = -\frac{i}{z\gamma}\kappa^{2}$ & $\omega = -\frac{i}{z\gamma}\kappa^{2}$ & $\omega = -\frac{i}{z\gamma}\kappa^{2}$ & $\omega = -\frac{i}{6z\gamma}\kappa^{2}$ \\ \hline
Shear (z $\ll$ 1)&$\omega=-\frac{i}{5\gamma}\kappa^{2}$ & $\omega=-\frac{i}{5\gamma}\kappa^{2}$ & $\omega=-\frac{i}{5\gamma}\kappa^{2}$ & $\omega=-\frac{i}{5\gamma}\kappa^{2}$ \\ \hline
Shear (z $\gg$ 1)&$\omega = -\frac{i}{z\gamma}\kappa^{2}$ & $\omega = -\frac{i}{z\gamma}\kappa^{2}$ & $\omega = -\frac{i}{z\gamma}\kappa^{2}$ & $\omega = -\frac{i}{z\gamma}\kappa^{2}$ \\ \hline
\end{tabular}
\caption{Comparison between this work and recent studies \cite{Hataei:2025mqf,Bajec:2025dqm} of dispersion relations in the long-wavelength limit for different collective modes. For completeness, the results of \cite{cercignani2002relativistic} are also shown.  \(z\ll 1 \) and \(z\gg 1\) correspond to the ultrarelativistic and nonrelativistic regimes, respectively.}
\end{table}

Here in this table, two distinct speeds of sound are given
\begin{align}
    c_{s}=
    \begin{cases}
        \frac{1}{3}, \ \ \ \ &z\ll1\\
        \sqrt{\frac{5}{3z}} , \ \ \ \ &z\gg1
    \end{cases}
     \ \ \ \ \ \ \ \ c_{s}^\prime=
    \begin{cases}
        \frac{1}{3}, \ \ \ \ &z\ll1\\
        \frac{1}{\sqrt{z}}, \ \ \ \ &z\gg1
    \end{cases}.
\end{align}
As seen from this table, it is evident that our work  provides complete and correct analytical expressions across all relevant limits. The calculations in \cite{Hataei:2025mqf} do not present a closed-form expression for the sound mode, and the authors explicitly note discrepancies between their result for the sound mode and previous findings. The only difference between our results and those of \cite{Bajec:2025dqm} lies in the calculation of the sound speed. Actually, the sound speed given in \cite{Bajec:2025dqm} is not derived from linear response theory, but rather obtained by  using the thermodynamic definition for the speed of sound. They specifically compute this for the case of zero chemical potential.  In other words, they are effectively computing the isothermal speed of sound, because in the absence of a chemical potential (i.e., at zero charge density), constant-entropy conditions reduce to constant-temperature conditions. This differs from the adiabatic (isentropic) speed of sound derived in our work.

Regarding on the comparison with \cite{cercignani2002relativistic}, we find that the ultrarelativistic limit of the imaginary part of the sound mode in  \cite{cercignani2002relativistic} does not agree with our results or with those reported in Refs.\cite{Bajec:2025dqm, Hu:2023elg, Romatschke:2015gic}.  
In addition, the nonrelativistic limit of the heat mode dispersion relation derived in \cite{cercignani2002relativistic} is inconsistent with all other existing studies listed above.

\subsection{Noncollective modes}
\label{more}

In the previous discussion, we considered the case where  \(\left ( c+\frac{\hat{p} }{\tau }\cdot l  \right ) \) is not zero, which corresponds to the collective modes. Now, we consider the complementary case
\begin{align}\label{none}
    \left ( c+\frac{\hat{p} }{\tau }\cdot l  \right )=c+\frac{\left | \mathbf{p}  \right | }{p^{0} } \cos\theta =0.
\end{align}
Thus it can be naturally interpreted as the sector of noncollective modes.

\begin{figure}[H]
    \centering
    \includegraphics[scale=0.73]{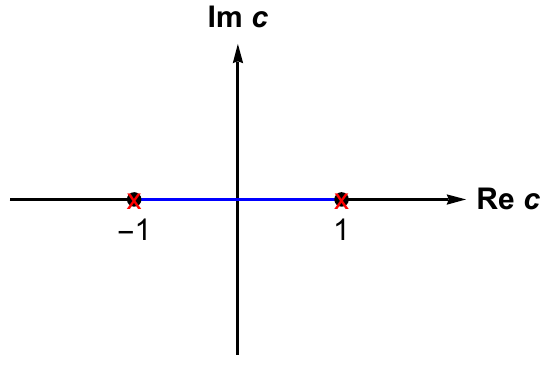}\ \ \ \ \ \ \ \ \ \ \ \ \ \ 
    \includegraphics[scale=0.73]{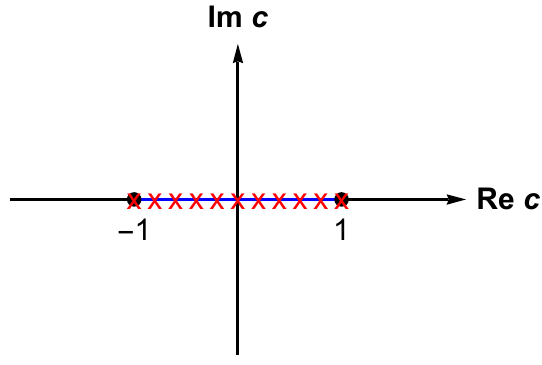}
    \caption{The nonanalytical structures contained in the secular equations. The blue lines represent the branch cuts of the massless (left panel) and massive (right panel) cases, while the red crosses represent branch points.}
    \label{fig:cut}
\end{figure}

Given the definition of $c$, we solve the above equation to obtain
\begin{align}\label{none_sol}
    \Re \omega=-\frac{|\mathbf{p}|\cos \theta}{p^{0}}\kappa
    ,\quad \Im \omega=-\gamma.
\end{align}
Strictly speaking, this expression does not constitute a dispersion relation in the conventional sense, even though it exhibits a relationship between frequency and wave number. The result explicitly contains the intermediate variables $p$ and $\cos\theta$, which must be integrated out [e.g. Eqs. \eqref{eq:rho} and \eqref{eq:G}] in the momentum integral to yield the final form of the dispersion relation.  It is convenient to start with Eqs.\eqref{Arho} and \eqref{Aij}  to clearly illustrate this point.
The coefficients $A_{ij}$ enter the secular equation derived from Eq.\eqref{eq:rho}, which determines the system’s normal mode structure. Since the retarded correlation functions are directly related to these normal modes, the analytic form of the correlators is inherently influenced by the structure of $A_{ij}$—in particular, its functional dependence on Eq.\eqref{eq:G} (see also the explicit forms of $A_{ij}$ given in Appendix \ref{appendix.c}). In the integral definition of Eq.\eqref{eq:G}, only the structure of the following form

\begin{align}\label{gcz}
    g\left(c,z\right)=
    \coth^{-1} \left ( c\frac{\sqrt{z^{2}+\mathbf{\left | \hat{p} \right | }^{2}   }}{\left | \mathbf{\hat{p}}  \right |}  \right ) 
    =\frac{1}{2}\ln \left( \frac{c\sqrt{z^{2}+\mathbf{\left | \hat{p} \right | }^{2}   }+\mathbf{\left | \hat{p} \right | }}
    { c\sqrt{z^{2}+\mathbf{\left | \hat{p} \right | }^{2}   }-\mathbf{\left | \hat{p} \right | }}\right)
\end{align}
contributes to the nonanalytic behavior. This expression naturally appears after performing the angular integration over $\theta$ in Eq.\eqref{Aij}. Consequently, the poles of the integrand in Eq.\eqref{Aij} explicitly displayed in Eqs.\eqref{none} and \eqref{none_sol} directly give rise to the branch cut encoded in the $\coth^{-1}$ structure within Eq.\eqref{eq:G}. It is sufficient to start with Eq.(\ref{gcz}) to analyze the singularity or nonanalytical structures within the secular equations $\Phi_1$ and $\Phi_2$. Evidently, $g\left(c,z\right)$ possesses the branch points.
When \(z\neq0\), the branch points vary with \(\left | \mathbf{\hat{p}} \right |\). Given a fixed \(\left | \mathbf{\hat{p}} \right |\), the branch points are determined by \(c=\pm \frac{\left |  \mathbf{\hat{p}} \right |}{\sqrt{z^{2}+\mathbf{\left | \hat{p} \right | }^{2}   }}\). Note that \(\left | \mathbf{\hat{p}} \right |\) ranges from $0$ to $\infty$, so we can obtain a set of branch points, i.e., \(\left \{ \pm \frac{\left |  \mathbf{\hat{p}} \right |}
{\sqrt{z^{2}+\mathbf{\left | p \right | }^{2}   }} \Big |
\left |  \mathbf{\hat{p}} \right |\in \left ( 0,\infty  \right ) \right \} \) \(=\left[-1,1\right]\), if interpreting the momentum integral as a summation over various $\mathbf{\hat{p}}$. All the branch points form a continuous branch cut line in the region \(\left[-1,1\right]\). Particularly, when \(z=0\), \(g\left(c,0\right)=\coth^{-1} c\), and the branch points coalesce into two end points, i.e., \(c=\pm 1\). As exhibited in Fig.\ref{fig:cut}, in the massless case (left panel), only two points, $c=\pm{1}$, serve as branch points, corresponding to $\Re \omega= \pm{\kappa}, \Im \omega=-\gamma$.
In order to define single-valued branches of the multivalued function, a branch cut connecting these two points must be introduced. However, the choice of branch cut is not unique. A common convention is to take the straight line segment from $-1$ to $1$ as the branch cut for  $\coth^{-1}c$.
When the mass is nonzero, every point along the interval from $-1$ to $1$ on the real axis becomes a branch point. Therefore, the most natural approach is to introduce a branch cut along this segment, connecting $-1$ to $1$ shown in the right panel.

\begin{figure}[H]
    \centering
    \includegraphics[scale=0.65]{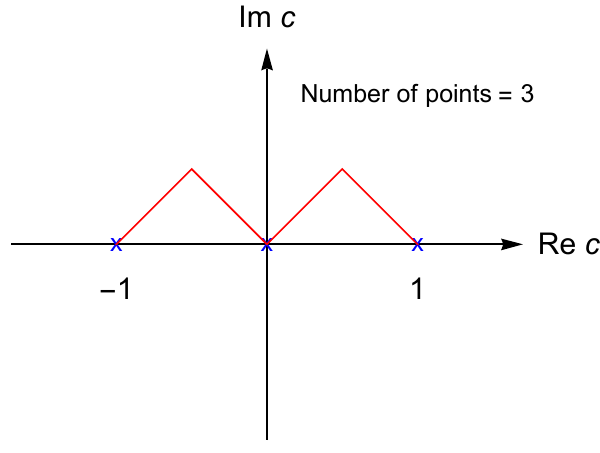}\ \ \ \ \ \ \ \ \ \ \ \ \ \ 
    \includegraphics[scale=0.65]{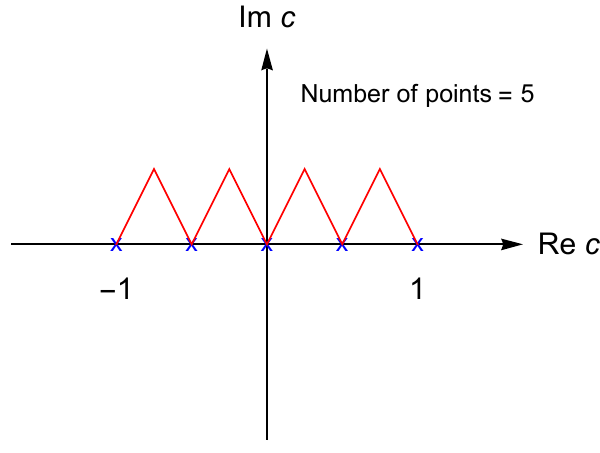}\ \ \ \ \ \ \ \ \ \ \ \ \ \ 
    \includegraphics[scale=0.65]{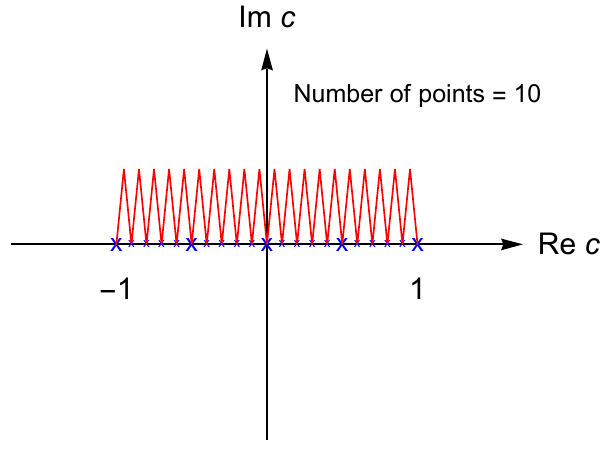}\ \ \ \ \ \ \ \ \ \ \ \ \ \ 
    \includegraphics[scale=0.65]{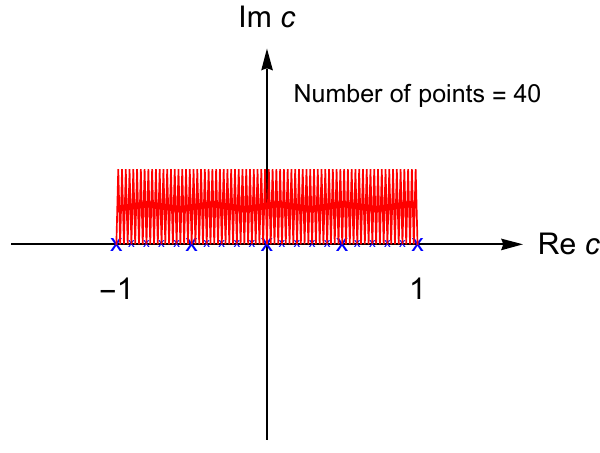}
    \caption{The branch cuts linking all branch points in the nonanalytic structure. The red line represents an arbitrarily deformable branch cut,  with the only requirement that it must pass through all branch points. Note that with increasing the point number, this one-dimensional line appears to densely fill a two-dimensional plane.} 
    \label{fig:nonanalysis}
\end{figure}

In \cite{Bajec:2025dqm}, the authors state that this nonanalytic structure corresponds to a nondeformable cut distinct from a conventional branch cut. When carrying out the momentum integral [e.g. Eqs. \eqref{eq:rho} and \eqref{eq:G}], deforming the integration contour will cut out a two-dimensional region of the complex $\omega$ plane where the correlator is not well defined [see Eq.(94) of \cite{Bajec:2025dqm} and the related discussion around it]. Different choices of integration contour cut out different regions of the complex plane that can be made arbitrarily large, which may be inconsistent with the general properties of the causal correlators. The only way to bypass the aforementioned problems is to fix the integration contour without deformation, which equivalently means that the branch cut along the segment connecting $-1$ to $1$ is uniquely fixed. Indeed, the present work aligns with this view: The distribution of singularities uniquely determines the shape and position of this cut line, leaving no freedom for deformation.  Since the entire line consists of branch points, the branch cut must connect all of them to properly define single-valued branches. Any deviation from the cut shown in Fig.\ref{fig:cut} would inevitably carve out a larger nonanalytical area compared to the horizontal line. In Fig.\ref{fig:nonanalysis}, the branch cut is depicted as the boundary of a two-dimensional region. As the number of branch points gradually approaches infinity, the resulting cut line visually converges to a continuous contour, effectively delineating an entire two-dimensional area including its interior. This reasoning is consistent with the argument presented in Ref.\cite{Bajec:2025dqm}. Nevertheless, it must be acknowledged that this argument has a weakness: Even in the massless case with only two branch points, mathematics allows for highly irregular deformations of the branch cut's path. As long as the paths do not intersect, one can, in principle, construct a cut that appears to densely cover a two-dimensional region—a procedure that is not forbidden by mathematical rigor. In this sense, we disagree with the criterion proposed in \cite{Bajec:2025dqm}, which uses the nondeformability of the branch cut as a mathematical means to distinguish between massive and massless cut structures.

For us, both in the massive and massless cases, there exists a physically motivated choice: The horizontal branch cut. This aligns with part of the argument presented in \cite{Brants:2024wrx}. First, the horizontal cut is the most natural choice: It carries the clearest physical interpretation and does not require any technical manipulations of the branch cut or integration over complex variables \cite{Brants:2024wrx}. Second, any deformation away from this choice leads to a larger nonanalytic region. Whether it is the two-dimensional “cut-out” region described in \cite{Bajec:2025dqm}, the boundary that visually appears to densely fill a two-dimensional area as discussed in this work, or even in the massless case, a highly deformed cut connecting just two end points that nearly covers a finite area (if deformations are allowed, arbitrarily complex ones should be permitted), the principle of minimality favors the horizontal cut as the simplest and most natural option. Finally, allowing arbitrary deformations away from the horizontal line would imply that one could always choose a contour such that hydrodynamic poles vanish from the principal Riemann sheet even if the wave vector is small enough, effectively obscuring the otherwise clear physical interpretation of these modes \cite{Romatschke:2015gic}. Therefore, the horizontal cut remains the most physically consistent and conceptually transparent choice. At this stage, we state that, from a physical standpoint, the horizontal line represents the most natural choice. In this picture, there is a sharp transition as one moves from the massless to the massive case - changing abruptly from two discrete branch points to a continuum of infinitely many.



These differing nonanalytic structures may have a direct impact on the radius of convergence of the hydrodynamic series. It is suggested that the radius of convergence (i.e., in the shear channel) is set by a collision of the hydrodynamic pole with a branch point \cite{Heller:2020hnq}. The transition from two branch points to a continuous distribution implies a qualitative change in the analytic structure, which is expected to influence the radius of convergence of the hydrodynamic expansion. Specifically, the hydrodynamic pole may collide with any branch point along the branch cut at finite mass, and this determines the radius of convergence of the hydrodynamic expansion. In contrast, within the massless RTA framework, such collisions can only occur between the hydrodynamic pole and the two branch points located at the end points of the cut \cite{Heller:2020hnq}.
This suggests that mass influences the system not only through conventional kinematical effects, but also by modifying the analytic structure of the correlation functions—an indirect yet significant mechanism. Understanding how different nonanalytic structures affect the convergence of hydrodynamic expansion in the massive case remains an important open question, which we plan to address in future work. Note as an aside,  the massive RTA yields a coupled structure between the sound and heat channels, in contrast to the massless RTA on which the analysis of hydrodynamic convergence is based \cite{Heller:2020hnq}.  As a result, not only the radius of convergence, but also the underlying mechanism may change significantly. For example, in Ref.\cite{Heller:2020hnq}, the radius of convergence in the sound channel is determined by the collision of two sound poles on a nonprincipal sheet of the Green’s function. In the massive case, however, this mechanism may be replaced by a collision between a hydrodynamic pole and a branch point, or even by a collision between the sound pole and the heat pole. Before a systematic and comprehensive analysis is given, we cannot rule out the possibility of a change in the collision mechanism. In other words, the massive RTA provides a tractable yet nontrivial framework that explicitly includes sound-heat channel coupling for studying the convergence of hydrodynamic dispersion relations in weakly coupled systems. We also hope to address this issue in the future.

The branch point singularity  is often related to a  finite radius of convergence. From the point of view of the quasinormal spectrum in holography, this point corresponds to the collision of the modes or level crossing, (see related discussion in this context \cite{Withers:2018srf,Grozdanov:2019kge}), while in kinetic theory, its origin is associated with the Landau damping. In a typical collisionless plasma, the ballistic propagator of free-streaming dynamics gives rise to the singularity of this type once the momentum integral is carried out. The physical picture underlying Landau damping, discovered by John Dawson, is the resonant interaction between waves and particles (the collective motion and noncollective motion) \cite{Dawson:1961}. 

   In the end of this section, we  compare our approach with the retarded correlator formalism and give several comments as follows: 
   \begin{itemize}
       \item In our presented linear mode analysis, the coupling between the sound and heat channels is explicitly considered.
       Within the framework of linear response theory, this coupling manifests through off-diagonal components of the retarded correlation functions, such as $G^R_{TJ}$ and $G^R_{JT}$, that describe the mutual influence of energy and charge fluctuations. Computing these cross-correlators requires evaluating the linear responses of both $\delta T^{\mu\nu}$ and $\delta J^\mu$ to external perturbations like $A_\alpha$ and $g_{\alpha\beta}$, which is typically more difficult (see Ref.\cite{Bajec:2024jez}). Our calculation provides a simple and convenient approach for studying linear response phenomena such as sound-heat coupling.  
       \item The branch point structure changes dramatically from two discrete points in the massless case to an infinite number of points in the massive case, even for arbitrarily small mass. This indicates that the transition is sudden and somewhat counterintuitive. It also reflects a limitation of our current method, as the secular equation we derive only captures the denominator of the retarded correlation function, while the numerator contains important contributions from the residues of poles and discontinuities across branch cuts. If one is solely concerned with extracting physical quantities such as the dispersion relations of normal modes, the present approach suffices. However, to fully understand the influence of poles and branch cuts on the real-time evolution of the system, it is essential to include the full structure of the retarded correlators, including its numerator. We argue that the numerator of the retarded correlation function may contain factors that regulate this abrupt transition \footnote{The explicit calculations given in \cite{Bajec:2025dqm} support the abrupt change in the case of $G^{2,2}_{JJ}$ as the
       denominator  of $G^{2,2}_{JJ}$ is trivial and analytic (see the notations and details in \cite{Bajec:2025dqm}).}, suggesting that a complete treatment at the level of the full retarded function could smooth out or modify the sharp change observed in the nonanalytic structure. It is worth noting that such a smoothing (if present) implies that, apart from the branch points at the end points retaining their weight, all other branch points lose their  weights. This bears a resemblance to the phenomenon of pole skipping in the context of quantum chaos \cite{Blake:2018leo,Grozdanov:2017ajz}, where the numerator and denominator of the correlation function simultaneously vanish at certain points.
   \end{itemize}

\section{SUMMARY AND OUTLOOK}
\label{outlook}
In this work, we carry out a comprehensive normal mode analysis of the linearized Boltzmann equation for massive particles within the relaxation time approximation. A key focus is placed on the emergence and behavior of collective excitations across different channels—specifically in the sound, heat, and shear channels. One notable observation is the coupling between the secular equations governing the dispersion relations of the sound and heat modes. This coupling is found to be intrinsically tied to the presence of a finite particle mass and disappears in the massless limit. 

To determine the conditions under which these collective modes exist, we apply the argument principle from complex analysis. Our results confirm that the onset transition behavior, previously identified in massless systems, persists in the case of massive transport, indicating a degree of universality in the nature of hydrodynamic mode emergence. By numerically solving for the critical wave number $\kappa_c$ at which each mode ceases to exist, we examine how this threshold varies with the scaled mass parameter. We find that the critical wave numbers for the heat and shear channels increase monotonically as the scaled mass increases. In contrast, the dependence of the sound channel’s $\kappa_c$ on the scaled mass reveals a nonmonotonic trend, suggesting a more intricate interplay between the system's mass and its linear response. 

Furthermore, we analytically derive the dispersion relations for all three channels in the long-wavelength regime. These expressions provide valuable insight into the low-momentum behavior of the collective modes and serve as benchmarks for future studies. Importantly, our analysis also incorporates collisionless dissipation effects known as Landau damping, which are encoded in the nonanalytic structure of the retarded correlation functions. We observe a striking difference in the branch cut structure between the massive and massless cases: While the massless system exhibits only two isolated branch points, the massive case features an infinite sequence of such points that coalesce into a continuous branch cut. This indicates a richer singularity structure in the complex plane when finite particle mass is taken into account. 

To conclude, this study sheds light on how the introduction of mass modifies the dynamics of collective and noncollective excitations in kinetic theory. It not only extends previous findings from the massless regime but also uncovers new phenomena such as the coupling of sound-heat channels and the qualitative change in the analytic structure of the response function. These results may have important implications for understanding the breakdown of hydrodynamics and the emergence of nonhydrodynamic behavior in systems with finite particle mass. Several  potential extensions for this research could be envisioned. First, as mentioned in the main text, the whole analysis can be straightforwardly generalized to nontrivial backgrounds, for instance, moving backgrounds or with external field. Second, it would be of great interest to examine whether the numerator of the other retarded correlation functions, other than $G_{JJ}^{2,2}$ mentioned in footnote 1,  allows the nonanalytic structures associated with Landau damping, i.e., the branch points and branch cuts, to smoothly approach their massless counterparts in the small-mass regime. This requires a comprehensive analysis of the full retarded correlators. Last but not least, calculations of higher-point functions based on kinetic theory, such as three-point functions \cite{Abbasi:2024pwz}, contain rich physical information. Extracting the physics of nonlinear responses from these calculations is also a highly interesting and promising direction for future study. We leave these investigations for future work.





\section*{Acknowledgments}

We thank Robbe Brants, Matej Bajec, and Alexander Soloviev for helpful comments and email communication. This work is supported by NSFC under Grants No.12505149 and No. 12405134.


\appendix

\section{ARGUMENT PRINCIPLE}\label{argu}

\begin{AP}
   
    \textup{ If f is analytic and nonzero on a rectifiable Jordan curve \(\eta\) 
    and meromorphic within \(\eta\), then}
    \begin{align}
         w\left ( f\left ( \eta  \right )  \right ) =
        \frac{1}{2\pi i}\int_{\eta }^{}  \frac{f'\left ( z \right ) }{f\left ( z \right ) } dz=N-P.
    \end{align}
\end{AP}
Here, \(N\) is the number of zeros inside \(\eta\) counted with multiplicity, and \(P\) is the number of poles inside \(\eta\) counted with multiplicity.

\begin{proof}
    If \(a\) is an n-fold zero inside \(\eta\),\(f\left(a\right)=\left(z-a\right)^{n}\varphi\left(a\right)\)
\ \ \ \ \ \(\varphi\left(a\right) \ne 0 \),
\begin{align}\label{ap1}
     \frac{f'}{f}=\frac{n(z-a)^{n-1} \varphi+(z-a)^{n} 
    \varphi'}{(z-a)^{n} \varphi}=\frac{n}{z-a}+\frac{\varphi^{\prime}}{\varphi}
    \ \ \ \ \ \ \text{Res}[\frac{f'}{f},a]=n.
\end{align}

Similarly, if \(b\) is an m-fold pole inside \(\eta\), \(\text{Res}[\frac{f'}{f},b]=-m\),
\begin{align}\label{ap2}
    \frac{1}{2\pi i}\int_{\eta }^{}  \frac{f'\left ( z \right ) }{f\left ( z \right ) } dz=N-P.
\end{align}
    
Compare Eqs.\eqref{ap1} and \eqref{ap2},
\begin{align}
     \frac{1}{2\pi i}\int_{\eta }^{}  \frac{f'\left ( z \right ) }{f\left ( z \right ) } dz
=\frac{1}{2\pi i}\int_{\eta }^{}  \frac {df\left ( z \right ) }{f\left ( z \right ) } =
\frac{1}{2\pi i}\triangle _{\eta } \ln f\left ( z \right ) =
\frac{1}{2\pi }\triangle _{\eta } f\left ( z \right )=
w\left ( f\left ( \eta  \right )  \right ) .
\end{align}
\end{proof}

\section{BICKLEY FUNCTION}
In the computation of the matrix elements appearing in the secular equation, we introduce a special function, the Bickley function, which is defined as follows: 
    \begin{align}
        Ki_{n}\left ( z \right ) \equiv 
         \int_{0}^{\infty } e^{-z\cosh x} 
        \frac{1   }{\cosh ^{n}x} dx\ \ \ \ \ \ \ \left(z > 0\right).
    \end{align}
Additionally, this function satisfies several useful identities:
\begin{align}
   Ki_{0}\left(z\right)&=K_{0}\left(z\right) \\
    Ki_{-n}\left(z\right)&    =\left(-1\right)^{n}\frac{d^{n}}{dz^{n}}K_{0}\left(z\right), \ \ \ \ \ \ \ \  n=1,2,3,..., \\
   Ki_{\alpha}\left(z\right)&=\frac{\sqrt{\pi}\Gamma\left(\frac{1}{z}\alpha\right)}
   {2\Gamma\left(\frac{1}{2}\alpha+\frac{1}{2}\right)},
   \ \ \ \ \ \ \ \  \alpha\neq0,-2,-4,....
\end{align}


\section{THE EXPLICIT FORM OF \(A_{ij}\)}
\label{appendix.c}
\begin{align}
A_{11}&=\hat{\gamma}(\frac{(zK_{2}+3K_{3})G(c,z,1,1)}{z((-1+z^{2})K_{2}^{2}+5zK_{2}K_{3}-z^{2}K_{3}^{2})}-\frac{(-K_{2}+zK_{3})G(c,z,2,1)}{z^{2}((-1+z^{2})K_{2}^{2}+5zK_{2}K_{3}-z^{2}K_{3}^{2})})-1,\\
A_{12}&=\hat{\gamma}(\frac{(K_{2}-zK_{3})G(c,z,1,1)/K_{2}-G(c,z,2,1)}{z^{2}\sqrt{((-1+z^{2})K_{2}^{2}+5zK_{2}K_{3}-z^{2}K_{3}^{2})}}),\\
A_{13}&=\hat{\gamma}  \left ( \frac{z^{2} \left ( zK_{3} -K_{2} \right ) }{\sqrt{z^{5} K_{2} K_{3} } }-\frac{cG\left ( c,z,2,1 \right )}{\sqrt{z^{5} K_{2} K_{3} }}\right ),\\
A_{21}&=\hat{\gamma}(\frac{K_{2}(zK_{2}+3K_{3})G(c,z,2,1)}{z((-1+z^{2})K_{2}^{2}+5zK_{2}K_{3}-z^{2}K_{3}^{2})^{\frac{3}{2}}}-\frac{K_{2}(-K_{2}+zK_{3})G(c,z,3,1)}{z^{2}((-1+z^{2})K_{2}^{2}+5zK_{2}K_{3}-z^{2}K_{3}^{2})^{\frac{3}{2}}}),\\
A_{22}&=\hat{\gamma}(\frac{K_{2}G(c,z,3,1)}{z^{2}((-1+z^{2})K_{2}^{2}+5zK_{2}K_{3}-z^{2}K_{3}^{2})}-\frac{(-K_{2}+zK_{3})G(c,z,2,1)}{z^{2}((-1+z^{2})K_{2}^{2}+5zK_{2}K_{3}-z^{2}K_{3}^{2})})-1,\\
A_{23}&=\hat{\gamma}(\frac{K_{2}(z^{4}K_{2}+3z^{3}K_{3})-cK_{2}G(c,z,3,1)}{z^{2}\sqrt{zK_{3}((-1+z^{2})K_{2}^{2}+5zK_{2}K_{3}-z^{2}K_{3}^{2})}}),\\
A_{31}&=\hat{\gamma}(\frac{c\sqrt{K_{2}}(-K_{2}+zK_{3})G(c,z,3,1)/\sqrt{zK_{3}}-c\sqrt{zK_{2}K_{3}}(zK_{2}+3K_{3})G(c,z,2,1)/K_{3}}{z^{2}((-1+z^{2})K_{2}^{2}+5zK_{2}K_{3}-z^{2}K_{3}^{2})}),\\
A_{32}&=\hat{\gamma}(\sqrt{5+\frac{(-1+z^{2})K_{2}}{zK_{3}}-\frac{zK_{3}}{K_{2}}}+\frac{zc\frac{K_{3}}{K_{2}}G(c,z,2,1)-c(G(c,z,2,1)+G(c,z,3,1))}{\sqrt{z^{5}\frac{K_{3}}{K_{2}}((-1+z^{2})K_{2}^{2}+5zK_{2}K_{3}-z^{2}K_{3}^{2})}}),\\
A_{33}&=\hat{\gamma}(c(-3-\frac{zK_{2}}{K_{3}})+\frac{c^{2}G(c,z,3,1)}{z^{3}K_{3}})-1.
\end{align}

\bibliographystyle{apsrev}  
\bibliography{references}{}   

\end{document}